\documentclass[
reprint,
superscriptaddress,
nofootinbib, aps]{revtex4-2}

\usepackage{physics}
\usepackage{bm}
\usepackage[T1]{fontenc}
\usepackage{amsmath,amsfonts,bbm, graphicx, color}
\usepackage{amssymb}
\usepackage{appendix}
\usepackage{mathtools}
\usepackage{dsfont}
\usepackage{float}
\usepackage{relsize}
\usepackage[dvipsnames]{xcolor}
\usepackage{scalerel}
\usepackage{tikz}
\usetikzlibrary{svg.path}
\usepackage{soul}
\usepackage{mathrsfs}
\usepackage{hyperref}
\usepackage{caption}
\usepackage{subcaption}
\hypersetup{
    colorlinks=true,
    linkcolor=blue,
    filecolor=magenta,      
    urlcolor=cyan,
    pdftitle={Overleaf Example},
    pdfpagemode=FullScreen,
    }

\DeclareMathAlphabet{\mathpzc}{OT1}{pzc}{m}{it}

\begin{document}

\title{Hawking radiation for detectors in superposition of locations outside a black hole}% Force line breaks with \\

\author{Jerzy Paczos}
\email{j.paczos@student.uw.edu.pl}
\affiliation{Faculty of Physics, University of Warsaw, ul. Pasteura 5, 02-093 Warszawa, Poland}
\author{Luis C.\ Barbado}
\email{luis.cortes.barbado@univie.ac.at}
% \author{\v{C}aslav Brukner}
% \email{caslav.brukner@univie.ac.at}
\affiliation{Institute for Quantum Optics and Quantum Information,
Austrian Academy of Sciences, Boltzmanngasse 3, 1090 Vienna, Austria}
\affiliation{Quantum Optics, Quantum Nanophysics and Quantum Information,
Faculty of Physics, University of Vienna, Boltzmanngasse 5, 1090 Vienna, Austria}
%\author{Andrzej~Dragan\orcid{0000-0002-5254-710X}}
%\email{dragan@fuw.edu.pl}
%\affiliation{Institute of Theoretical Physics, University of Warsaw, Pasteura 5, 02-093 Warsaw, Poland}
%\affiliation{Centre for Quantum Technologies, National University of Singapore, 3 Science Drive 2, 117543 Singapore, Singapore}

\date{\today}% It is always \today, today,

\begin{abstract}
Hawking radiation is the proposed thermal black-body radiation of quantum nature emitted from a black hole. One common way to give an account of Hawking radiation is to consider a detector that follows a static trajectory in the vicinity of a black hole and interacts with the quantum field of the radiation. In the present work, we study the Hawking radiation perceived by a detector that follows a quantum superposition of static trajectories in Schwarzschild spacetime, instead of a unique well-defined trajectory. We analyze the quantum state of the detector after the interaction with a massless real scalar field. We find that for certain trajectories and excitation levels, there are non-vanishing coherences in the final state of the detector. We then examine the dependence of these coherences on the trajectories followed by the detector and relate them to the distinguishability of the different possible states in which the field is left after the excitation of the detector. We interpret our results in terms of the spatial distribution and propagation of particles of the quantum field.
\end{abstract}

\maketitle

%--------------------------------------------
\section{Introduction}
%--------------------------------------------
Hawking radiation \cite{Hawking1, Hawking2} is arguably the most important result of quantum field theory in curved spacetimes (QFT-CS). Hawking's prediction implies that there is some thermal black-body radiation released outside a black hole's horizon. It is a striking quantum phenomenon since, according to classical general relativity, nothing can escape from inside a black hole. This has momentous consequences such as the gradual evaporation of black holes and their eventual disappearance. According to theoretical predictions, Hawking radiation is very weak --- a black hole of one solar mass would emit radiation like a black body with a temperature of about 60 nanokelvins. Due to the extraordinary weakness of the effect, it has not been found in astronomical observations. Experimental evidence has been reported in black hole analogs employing Bose-Einstein condensates \cite{Barcelo, Steinhauer, Nova, Kolobov}.

A common tool used in QFT-CS, and in particular in the study of the properties of Hawking radiation, is a particle detector. A simple model of a detector as a two-state quantum system coupled to the quantum field was first proposed by Unruh \cite{Unruh} and DeWitt \cite{DeWitt}, and since then it has been widely used to investigate the properties of Hawking radiation in various spacetimes \cite{Hartle, Israel, Hodgkinson1, Hodgkinson2, Gibbons}. This model has been extensively used to investigate also the ``flat spacetime relative of Hawking radiation'', namely the Unruh effect \cite{Unruh, DeWitt, Letaw, Grove, Barbado, Zych, Bruschi}, which is an effect experienced by accelerated observers moving in flat spacetime: the vacuum state for an inertial observer is perceived by the accelerated one as a thermal bath with temperature proportional to the acceleration.

Recently, there has been some interest in the relativistic effects perceived by particle detectors following a superposition of trajectories \cite{Grochowski, Paczos, Debski, Zych, Barbado}. In particular, in \cite{Zych, Barbado} the Unruh effect for detectors in a superposition of classical accelerated trajectories is studied. The authors of \cite{Zych} analyze the response of the Unruh-DeWitt detector and discover the presence of novel interference dynamics in its emission and absorption spectra. On the other hand, the work \cite{Barbado} is focused on the coherences left in the state of the detector after the interaction with the field. It is shown there that the excitation of the particle detector due to the Unruh effect does not need to imply complete decoherence in its final state.

Building on the idea of \cite{Barbado}, and making extensive use of the technical results from \cite{Hodgkinson1}, in this work we study Hawking radiation as perceived by a particle detector following a quantum superposition of classical static trajectories in the vicinity of the Schwarzschild black hole. We derive the formula for the final state of the detector after its interaction with the massless scalar quantum field of the radiation and analyze the coherences left in it. We consider two specific initial states of the field, namely, Hartle-Hawking and Unruh state. In both cases, we find that coherences in the final state of the detector are in general present, in an analogous way as for the Unruh effect in \cite{Barbado}. We derive the conditions which must be satisfied in order to have non-zero coherence after the excitation of the detector and analyze the dependence of the coherences on the trajectories followed by the detector, and the energies of the excitation. As part of our considerations, we discuss the state in which the field is left after the excitation of the detector. This is closely related to the presence of coherences in the final state of the detector --- these coherences have their origin in the non-distinguishability of the final states of the field left along different trajectories. Ultimately, based on this discussion, we draw conclusions concerning the spatial shape and propagation of the particles corresponding to the scalar field.

This article is organized as follows. In Sec.~\ref{sec: statement} we set up the problem, introducing the metric, the field, the detector model, the trajectories, and the interaction. In Sec.~\ref{sec: results} we give the results obtained for the state of the detector after the interaction. The results are visualized and interpreted in Sec.~\ref{sec: interpretation}. Finally, in Sec.~\ref{sec: conclusions} we draw some conclusions and identify possible directions for future developments of the work. In Appendix~\ref{Appendix A} we present the detailed calculations leading to the results in Sections \ref{sec: results} and \ref{sec: interpretation}.

%--------------------------------------------
\section{Statement of the problem}\label{sec: statement}
%--------------------------------------------

Throughout the article, we will use natural units $\hbar=c=G=k_\text{B}=1$. Let us consider a real scalar massless quantum field $\hat{\phi}(t,r,\theta,\varphi)$ on a Schwarzschild black hole
\begin{equation}\label{schwarzschild metric}
\begin{split}
\mathrm{d}s^2=&-\left(1-\frac{2 M}{r}\right)\mathrm{d}t^2+\left(1-\frac{2 M}{r}\right)^{-1}\mathrm{d}r^2\\
&+r^2\left(\mathrm{d}\theta^2+\sin^2\theta\mathrm{d}\varphi^2\right),
\end{split}
\end{equation}
where $M$ stands for the mass of the black hole. We consider a pointlike detector that couples to the field. The detector has several internal excitation levels $\{\ket{0}_\text{D},\ket{\omega_1}_\text{D},\ket{\omega_2}_\text{D},\ldots\}$, with energies $0<\omega_1<\omega_2<\ldots$, and an external degree of freedom corresponding to its trajectory. We will consider quantum superpositions of static trajectories ($r,\theta,\varphi=\text{const}.$), described by the Hilbert space spanned by the states $\{\ket{1}_\text{T},\ket{2}_\text{T},\ldots\}$. The states $\ket{n}_\text{T}$ correspond to well-defined static trajectories, and are defined by the relation
\begin{equation}\label{trajectory states}
(\hat{t}(\tau),\hat{r}(\tau),\hat{\theta}(\tau),\hat{\varphi}(\tau))\ket{n}_\text{T}=\left(\alpha_n\tau,r_n,\theta_n,\varphi_n\right)\ket{n}_\text{T},
\end{equation}
where $\tau$ is the proper time of the detector, $\alpha_n$ stands for the gravitational time dilation factor
\begin{equation}
\alpha_n=\frac{1}{\sqrt{1-2M/r_n}},
\end{equation}
and $r_n$, $\theta_n$, and $\varphi_n$ are constants. We assume that all trajectories in the basis are fully distinguishable, $\braket{n}{m}_\text{T}=\delta_{nm}$. We will denote the operator standing on the LHS of \eqref{trajectory states} by $\hat{x}(\tau)$ --- it is the operator associating Schwarzschild coordinates to each trajectory state.

The coupling of the detector and the scalar field is given by the following interaction term in the Hamiltonian:
\begin{equation}\label{interaction hamiltonian}
    \hat{H}_\text{I}(\tau)=-\varepsilon\chi(\tau)\hat{m}(\tau)\hat{\phi}(\hat{x}(\tau)),
\end{equation}
where $\varepsilon\ll1$ is a weak coupling constant, $\chi(\tau)$ is a switching function that controls the intensity of the coupling, and $\hat{m}(\tau)$ is the monopole moment of the detector.

For the switching function, we choose the square root of a Gaussian function
\begin{equation}\label{switching function}
    \chi(\tau)=\frac{1}{(2\pi)^{1/4}}\mathrm{e}^{-\tau^2/(4T^2)},
\end{equation}
with $T$ being the approximate time duration of the interaction. We assume that
\begin{equation}\label{T assumption}
T\sim\frac{1}{\varepsilon\omega_1}\gg \frac{1}{\omega_1}\ge\frac{1}{\omega_i},
\end{equation}
which means that we switch the interaction adiabatically. The monopole moment equals
\begin{equation}
    \hat{m}(\tau)=\sum_i\zeta_i\mathrm{e}^{\mathrm{i}\omega_i\tau}\ket{\omega_i}\bra{0}_\text{D}+\text{h.c.},
\end{equation}
where $\zeta_i$ is the coupling amplitude from the ground state $\ket{0}_\text{D}$ to the excited state $\ket{\omega_i}_\text{D}$.

We consider the initial state of the system to be
\begin{equation}
    \ket{\Psi(\tau\to-\infty)}=\ket{0}_\text{D}\ket{\Omega}_\text{F}\left(\sum_n A_n\ket{n}_\text{T}\right),
\end{equation}
where $A_n$ is the amplitude for the trajectory $\ket{n}_\text{T}$, and $\ket{\Omega}_\text{F}$ is the initial state of the field (Unruh or Hartle-Hawking state).

Up to the first order in $\varepsilon$, the state of the system after the interaction is given by
\begin{equation}
    \ket{\Psi(\tau\to\infty)}=\left(\hat{\mathrm{I}}-\mathrm{i}\int_{-\infty}^\infty\mathrm{d}\tau\hat{H}_\text{I}(\tau)\right)\ket{\Psi(\tau\to-\infty)}.
\end{equation}
This can be rewritten in a generic way
\begin{equation}
\begin{split}
    \ket{\Psi(\tau\to\infty)}=&\ket{0}_\text{D}\ket{\Omega}_\text{F}\left(\sum_n A_n\ket{n}_\text{T}\right)\\
    &+\mathrm{i}\varepsilon\sum_{i,n}\zeta_i A_n\ket{\omega_i}_\text{D}\ket{\omega_i,n}_\text{F}\ket{n}_\text{T},
\end{split}
\end{equation}
where
\begin{equation}\label{field state}
\begin{split}
    \ket{\omega_i,n}_\text{F}:=&(\mathrm{i}\varepsilon\zeta_i A_n)^{-1}\bra{\omega_i}_\text{D}\bra{n}_\text{T}\ket{\Psi(\tau\to\infty)}\\
    =&-(\varepsilon\zeta_i)^{-1}\bra{\omega_i}_\text{D}\bra{n}_\text{T}\int_{-\infty}^\infty\mathrm{d}\tau\hat{H}_\text{I}(\tau)\ket{0}_\text{D}\ket{\Omega}_\text{F}\ket{n}_\text{T}
\end{split}
\end{equation}
is the state of the field left after the interaction with the atom following the trajectory $\ket{n}_\text{T}$ causing its excitation to the level $\ket{\omega_i}_\text{D}$.
%\luis{[I'd rather say ``In order to compute the state of the detector, we trace for the field, or something like that. We will comment on the state of the field, so it is not that true that we're not interested.]}

%--------------------------------------------
\section{Results}\label{sec: results}
%--------------------------------------------
In order to compute the final state of the detector, we trace out the field degrees of freedom:
\begin{widetext}
\begin{equation}\label{detector-trajectory state}
\begin{split}        \rho_\text{DT}:=&\Tr_\text{F}\left(\ket{\Psi(\tau\to\infty)}\bra{\Psi(\tau\to\infty)}\right)\\
=&\left(\sum_{n,m}A_n^*A_m\ket{m}\bra{n}_\text{T}\right)\ket{0}\bra{0}_\text{D}+\varepsilon^2\sum_{i,j,n,m}\zeta_i^*A_n^*\zeta_j A_m\braket{\omega_i,n}{\omega_j,m}_\text{F}\ket{\omega_j}\bra{\omega_i}_\text{D}\ket{m}\bra{n}_\text{T}.
\end{split}
\end{equation}
The quantities that remain to be computed in the above formula are the scalar products $\braket{\omega_i,n}{\omega_j,m}_\text{F}$. They are computed in Appendix~\ref{Appendix A}. The result is
\begin{equation}\label{detector-trajectory state 2}
\begin{split} 
\rho_\text{DT}\approx&\left(\sum_{n,m}A_n^*A_m\ket{m}\bra{n}_\text{T}\right)\ket{0}\bra{0}_\text{D}+\frac{\varepsilon^2T}{2\pi}\bigg[\sum_n|A_n|^2\ket{n}\bra{n}_\text{T}\sum_i|\zeta_i|^2\sigma_{in}\frac{\omega_i}{\mathrm{e}^{q_{in}/T_\text{H}}-1}\ket{\omega_i}\bra{\omega_i}_\text{D}\\&+\sum_{\substack{n,m\\n\neq m}}A_n^*A_m\ket{m}\bra{n}_\text{T}\sum_{\substack{i,j\\i\neq j}}^{\text{cond}}\zeta_i^*\zeta_j \Lambda_{nm}^{ij}\sqrt{\sigma_{in}\sigma_{jm}}\frac{\sqrt{\omega_i\omega_j}}{\mathrm{e}^{q_{in}/T_\text{H}}-1}\ket{\omega_j}\bra{\omega_i}_\text{D}\bigg].
\end{split}
\end{equation}
\end{widetext}

Let us explain the notation in \eqref{detector-trajectory state 2}. First, $\sigma_{in}=\sigma(\omega_i,r_n)$ accounts for correction to the radiation due to the radial position of the trajectory, including dispersion and backscattering of the radiation, and also depending on the concrete initial state of the field. We will comment on this in the interpretation section.

Next, we introduce
\begin{equation}
    q_{in}:=\frac{\omega_i}{\alpha_n},
\end{equation}
which is the energy of the $i$th level rescaled by the blueshift factor at the trajectory $\ket{n}_\text{T}$, and
\begin{equation}\label{Hawking temperature}
    T_\text{H}:=\frac{1}{8\pi M},
\end{equation}
which stands for the Hawking temperature of the black hole radiation. In the last sum, we use the label ``cond'', which denotes that we sum only over the terms for which
\begin{equation}\label{condition}
    q_{in}\approx q_{jm}
\end{equation}
holds to order $\varepsilon$. Finally,
\begin{equation}
    \Lambda_{nm}^{ij}:=\frac{\braket{\omega_i,n}{\omega_j,m}_\text{F}}{\sqrt{\braket{\omega_i,n}{\omega_i,n}_\text{F}\braket{\omega_j,m}{\omega_j,m}_\text{F}}}
\end{equation}
is the inner product between the normalized states of the field. In Appendix~\ref{Appendix A} one can find the explicit formulae for the factors $\sigma_{in}$ and $\Lambda_{nm}^{ij}$ together with the derivations. In particular, it is shown there that the product $\braket{\omega_i,n}{\omega_j,m}_\text{F}$ vanishes unless the condition \eqref{condition} is satisfied, which justifies the constraint on the last sum in \eqref{detector-trajectory state 2}.

It is also interesting to consider the state of the internal energy levels left after performing the measurement of the trajectory, and finding it to be e.g., $\ket{\eta}_\text{T}=\sum_n B_n \ket{n}$. Up to a normalization constant, such a state is given by
\begin{widetext}
\begin{equation}\label{detector state}
\begin{split}
    \rho^\text{measure}_\text{D}:=\Tr_\text{T}(\ket{\eta}\bra{\eta}_\text{T}\rho_\text{DT})\approx&\left(\sum_{n,m}A_n^*B_m^*A_mB_n\right)\ket{0}\bra{0}_\text{D}+\frac{\varepsilon^2T}{2\pi}\bigg[\sum_n|A_nB_n|^2\sum_i|\zeta_i|^2\sigma_{in}\frac{\omega_i}{\mathrm{e}^{q_{in}/T_\text{H}}-1}\ket{\omega_i}\bra{\omega_i}_\text{D}\\&+\sum_{\substack{n,m\\n\neq m}}A_n^*B_m^*A_mB_n\sum_{\substack{i,j\\i\neq j}}^{\text{cond}}\zeta_i^*\zeta_j \Lambda_{nm}^{ij}\sqrt{\sigma_{in}\sigma_{jm}}\frac{\sqrt{\omega_i\omega_j}}{\mathrm{e}^{q_{in}/T_\text{H}}-1}\ket{\omega_j}\bra{\omega_i}_\text{D}\bigg].
\end{split}
\end{equation}
\end{widetext}
The main conclusion we can already extract is that any time the product $\braket{\omega_i,n}{\omega_j,m}_\text{F}$ does not vanish for different excited states and different trajectories, the internal state of the detector is not just a classical mixture of spectra perceived along the different trajectories, but rather some coherences remain between the different excited states.

Formulae \eqref{detector-trajectory state 2} and \eqref{detector state} constitute the main results of our work. In the following section, we will discuss them physically.

\section{Interpretation of the results}\label{sec: interpretation}
%--------------------------------------------
Let us comment on the different terms appearing in \eqref{detector-trajectory state 2} in detail. The first term is the zeroth order in $\varepsilon$ and corresponds to the case in which interaction between the detector and the field yielded no excitation. The terms with the factor $\varepsilon^2 T$ are first order in $\varepsilon$ [recall that we assumed in \eqref{T assumption} that $T\sim\varepsilon^{-1}$] and they correspond to the contribution of the interaction. There are both diagonal and off-diagonal terms contained in the sum.

The diagonal terms for each trajectory $n$ follow a Planckian probability distribution with the Hawking temperature $T_\text{H}$, rescaled by a position dependent factor $\sigma_{in}$ and filtered by the coupling amplitudes $\zeta_i$ for each frequency. These are the contributions of the radiation of the field for each trajectory separately, combined in an incoherent way. Therefore, our construction reproduces the standard black hole radiation effect for detectors following a well-defined classical trajectory.

The most relevant result is the appearance of the off-diagonal terms \cite{Barbado}, corresponding to coherences between different trajectories. The necessary condition for these terms to appear is given by \eqref{condition}. Physically, this condition means that the quotients of the energies $\omega_i$ and $\omega_j$ with the blueshift factors $\alpha_n$ and $\alpha_m$, respectively, are (approximately) the same in both trajectories and excitations compared. This means that the detector's two excited states must degenerate in energy as perceived by any static observer. If the condition is satisfied, the corresponding off-diagonal term is the square root of the product of the Planckian spectra for the two corresponding trajectories and excited frequencies rescaled by the greybody factors and weighted with the product $\Lambda_{nm}^{ij}$.

These coherences arise from the properties of the field state $\ket{\omega_i,n}_\text{F}$ after the excitation of the detector. The perturbations left on the field that correspond to transitions to different energy levels of the detector and different trajectories are not always fully distinguishable. When these perturbations overlap, the product $\Lambda_{nm}^{ij}$ is nonzero, and the off-diagonal terms appear. This occurs because, when the compared states of the field are not fully distinguishable, no full entanglement is created between the excited states of the detector and the field due to the interaction. Therefore, tracing out the field does not introduce full decoherence in the state of the detector.

Provided that the condition \eqref{condition} is satisfied, the value of the product depends only on the radial distances $r_n$ and $r_m$ of the trajectories, the angle $\theta$ between them, and the value of $q_{in}=\omega_i/\alpha_n$. In the following subsections, we will separately discuss the functional dependence of $\sigma_{in}$ and $\Lambda_{nm}^{ij}$ for Hartle-Hawking and Unruh states.
%--------------------------------------------
\subsection{Hartle-Hawking state}
\begin{figure}
    \centering
    \includegraphics[width=0.4\textwidth]{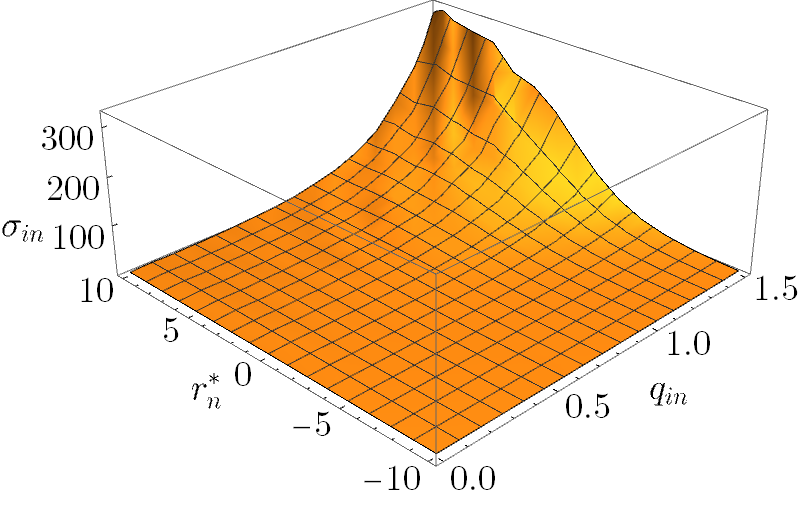}
    \caption{The value of the $\sigma_{in}$ for the Hartle-Hawking state as a function of the tortoise coordinate $r_n^*$ and the energy of the excitation rescaled by appropriate blueshift factor $q_{in}$.}
    \label{fig::sigma hartle-hawking}
\end{figure}

%--------------------------------------------
Let us begin by considering the Hartle-Hawking state. First, we visualize in Fig.~\ref{fig::sigma hartle-hawking} the dependence of $\sigma_{in}$ on the radial distance $r_n$ of the trajectory $\ket{n}_\text{T}$ and on the energy of the excitation $\omega_i$. We plot the value of $\sigma_{in}$ as a function of the tortoise coordinate $r_n^*=r_n+2M\ln\left(\frac{r_n}{2M}-1\right)$, corresponding to the trajectory $n$, and the energy of the excitation rescaled by the blueshift factor $q_{in}=\omega_i/\alpha_n$. We see that the value of $\sigma_{in}$ grows both with $r_n^*$ and $q_{in}$. %However, while it tends to some finite value for growing $r_n^*$, it seems to grow unboundedly with $q_{in}$.

\begin{figure}
\centering
\begin{subfigure}{0.4\textwidth}
         \centering
         \includegraphics[width=\textwidth]{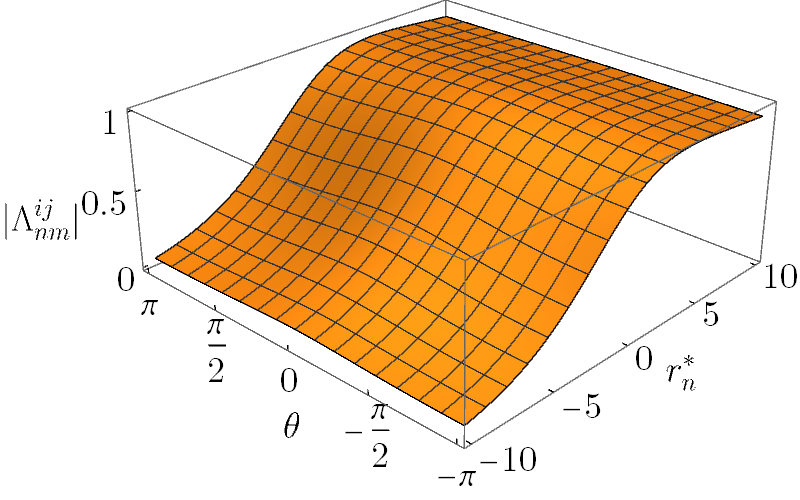}
         \caption{$q_{in}=0.01$}
         \label{fig::hawking product (a)}
\end{subfigure}
\hfill
\begin{subfigure}{0.4\textwidth}
         \centering
         \includegraphics[width=\textwidth]{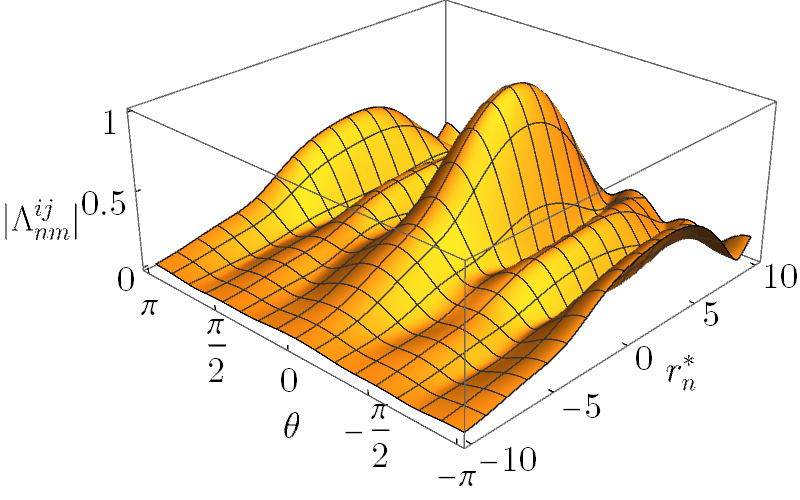}
         \caption{$q_{in}=0.5$}
         \label{fig::hawking product (b)}
\end{subfigure}
\begin{subfigure}{0.4\textwidth}
         \centering
         \includegraphics[width=\textwidth]{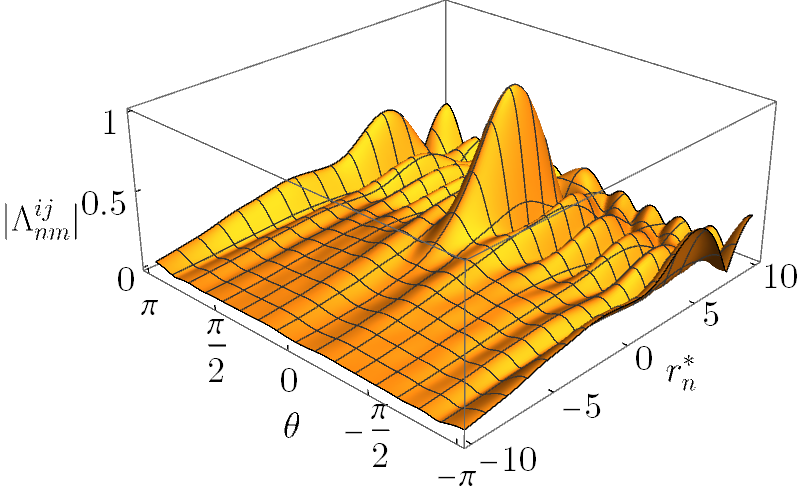}
         \caption{$q_{in}=1$}
         \label{fig::hawking product (c)}
\end{subfigure}
\hfill
\begin{subfigure}{0.4\textwidth}
         \centering
         \includegraphics[width=\textwidth]{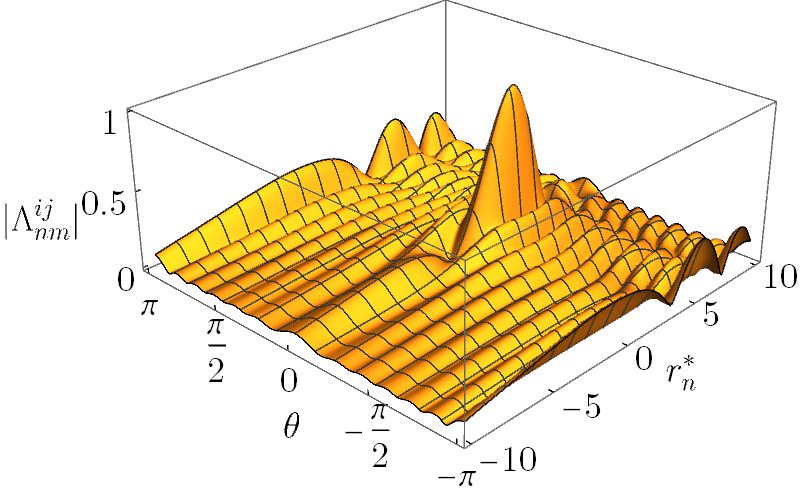}
         \caption{$q_{in}=1.5$}
         \label{fig::hawking product (d)}
\end{subfigure}
\caption{The absolute value of the normalized product $|\Lambda_{nm}^{ij}|$, for the initial state of the field being the Hartle-Hawking state, as a function of the tortoise coordinate $r_n^*$ corresponding to the trajectory $\ket{n}_\text{T}$, and the angle $\theta$ formed by the trajectories $\ket{n}_\text{T}$ and $\ket{m}_\text{T}$. The radial coordinate $r_m^*$ corresponding to $\ket{m}_\text{T}$ is fixed to be $r_m^*=4M$.}
\label{fig::hawking product}
\end{figure}

Now, let us analyze the functional dependence of the factors $\Lambda_{nm}^{ij}$ on the trajectories $n$ and $m$. In Fig.~\ref{fig::hawking product} we plot the absolute value $|\Lambda_{nm}^{ij}|$ for fixed tortoise coordinate $r_m^*=4M$ corresponding to the trajectory $m$, and variables the tortoise coordinate $r_n^*$ corresponding to the trajectory $n$, and angle $\theta$ between the trajectories. The maximum value of 1 is reached at only one point, $r_n^*=4M$ and $\theta=0$, corresponding to the coincidence of the trajectories and, because of the condition \eqref{condition}, the coincidence of the excited states. For $q$ large enough, it decays quickly in all directions with oscillatory behavior in $r_n^*$ and $\theta$. The oscillatory behavior becomes more prominent as the value of $q_{in}$ increases. Let us highlight the relatively large coherence on the antipodes appearing for larger values of $q_{in}$.

\begin{figure*}
\centering
\begin{subfigure}{0.45\textwidth}
         \centering
         \includegraphics[width=\textwidth]{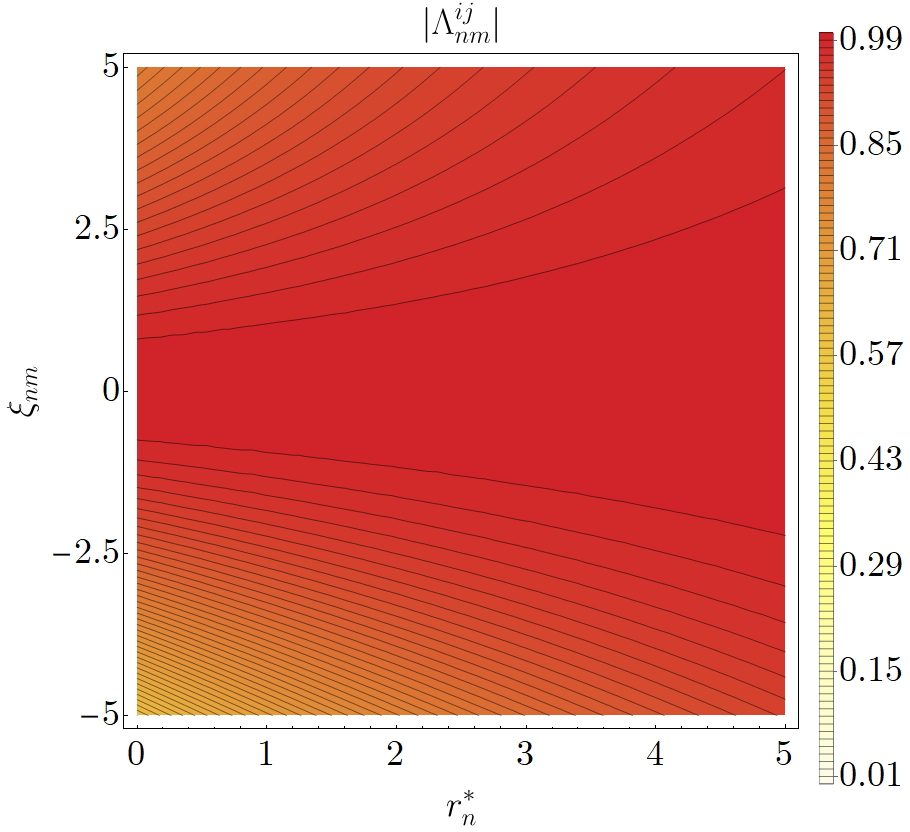}
         \caption{$q_{in}=0.01$}
         \label{fig::hawking rr product (a)}
\end{subfigure}
\hfill
\begin{subfigure}{0.45\textwidth}
         \centering
         \includegraphics[width=\textwidth]{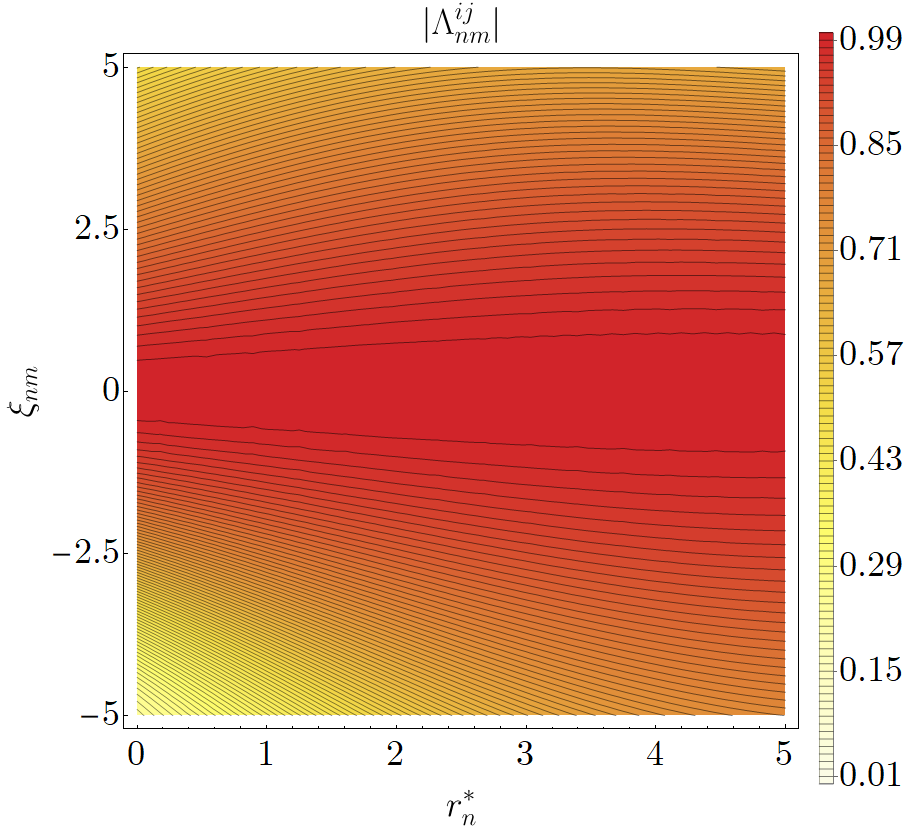}
         \caption{$q_{in}=0.5$}
         \label{fig::hawking rr product (b)}
\end{subfigure}
\hfill
\begin{subfigure}{0.45\textwidth}
         \centering
         \includegraphics[width=\textwidth]{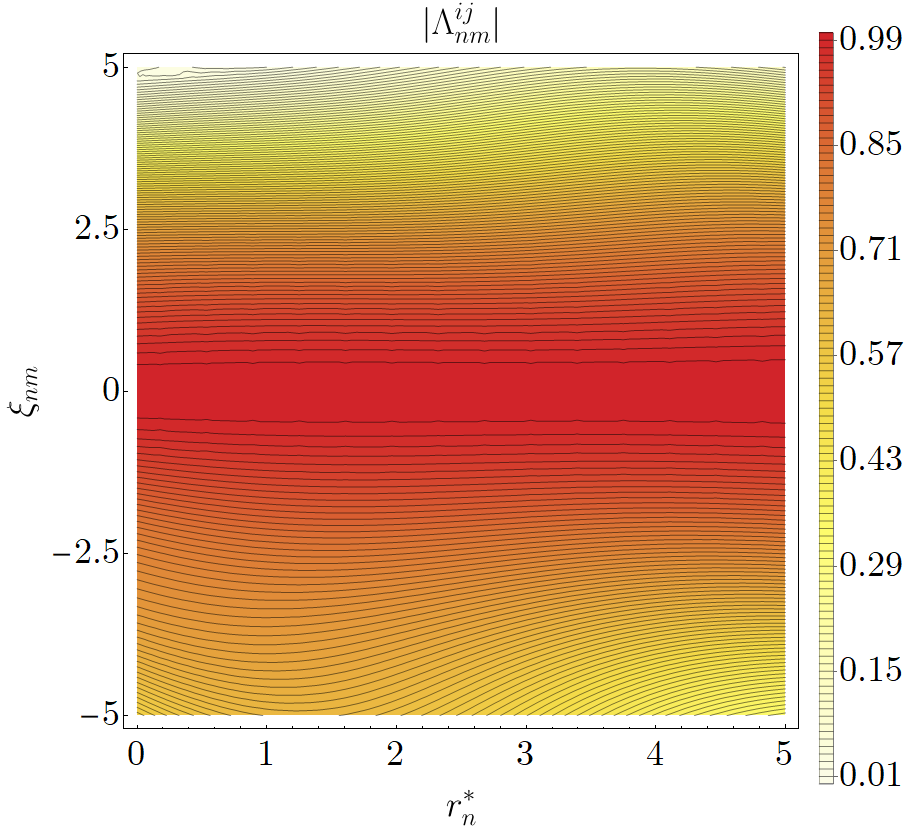}
         \caption{$q_{in}=1$}
         \label{fig::hawking rr product (c)}
\end{subfigure}
\hfill
\begin{subfigure}{0.45\textwidth}
         \centering
         \includegraphics[width=\textwidth]{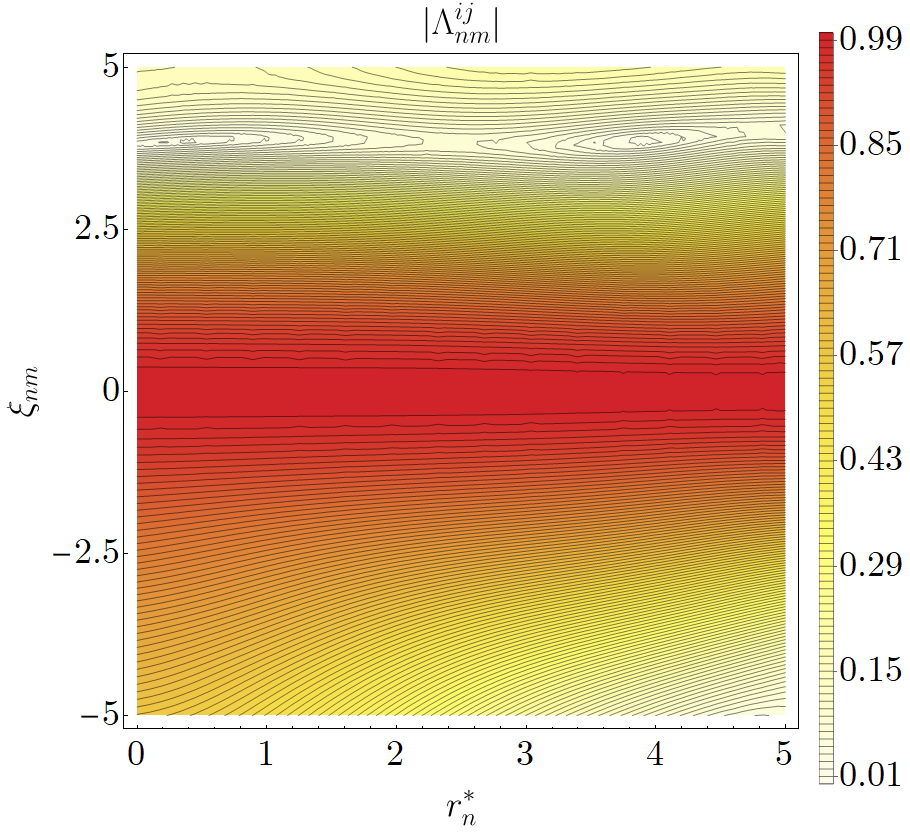}
         \caption{$q_{in}=1.5$}
         \label{fig::hawking rr product (d)}
\end{subfigure}
\caption{The absolute value of the normalized product $|\Lambda_{nm}^{ij}|$, for the initial state of the field being the Hartle-Hawking state, as a function of the tortoise coordinate $r_n^*$ corresponding to the trajectory $\ket{n}_\text{T}$, and the radar distance $\xi_{nm}$ between trajectories $\ket{n}_\text{T}$ and $\ket{m}_\text{T}$. The angle between the trajectories is fixed to be $\theta=0$.}
\label{fig::hawking rr product}
\end{figure*}

The plots in Fig.~\ref{fig::hawking rr product} show how the value of $\Lambda_{nm}^{ij}$ changes when we fix the angle between the trajectories $\theta=0$ and variate the radial positions of both trajectories. We make the plots of $|\Lambda_{nm}^{ij}|$ as a function of the tortoise coordinate $r_n^*$ and a radar distance $\xi_{nm}=2(r_m^*-r_n^*)/\alpha_n$. The meaning of the radar distance $\xi_{nm}$ is the following: it is the time in which light travels from trajectory $n$ to trajectory $m$ back and forth, measured by an observer following the trajectory $n$. We see that the behavior of $\Lambda_{nm}^{ij}$ is asymmetric in $\xi_{nm}$, although for larger values of $q_{in}$ and small values of $\xi_{nm}$ the difference between in and out directions becomes less prominent. Moreover, if we consider $q_{in}$ large enough, we observe oscillations of the value of $|\Lambda_{nm}^{ij}|$ when we range away from the horizon.
%

%Now, for $q$ significantly larger than 0, the value of $\Lambda_{nm}^{ij}$ is very sensitive to the relative distance $r_n^*-r_m^*$ for large $r_n^*$ and $r_m^*$ --- even for a small difference between $r_n^*$ and $r_m^*$ the value of $|\Lambda_{nm}^{ij}|$ drops rapidly. Interestingly, the growing value of $q$ enlarges the range of distances near the horizon where larger differences $r_n^*-r_m^*$ do not imply a significant drop of $|\Lambda_{nm}^{ij}|$. It is worth noting that for $q$ close to 0, the plots for the Unruh and the Hartle-Hawking state of the field are very similar.

Let us discuss the physical interpretation of the results for the Hartle-Hawking state. This state is perceived by a static observer as a thermal bath of particles, and the excitation of the detector is caused (according to such an observer) by the absorption of a particle from the bath. In this sense, the situation is analogous to the one considered in \cite{Barbado}, where the Unruh effect for detectors in a superposition of accelerations was analyzed. First, the long time of the interaction between the field and the detector implies that the dispersion in the energy of the excitation must be tiny, which is expressed by the condition \eqref{condition}. Secondly, the dependence of $|\Lambda_{nm}^{ij}|$ on $\theta$ and $r_n^*$ for fixed $r_m^*$ (Fig.~\ref{fig::hawking product}) can be interpreted as providing a notion of spatial localization of the absorbed particle. We see that the particles are completely delocalized for small values of $q_{in}$, but become well-localized with growing $q_{in}$. In view of this interpretation, the peaks of $|\Lambda_{nm}^{ij}|$ on the antipodes are particularly interesting, although we do not have any particular explanation for such phenomenon.
%--------------------------------------------
\subsection{Unruh state}
%--------------------------------------------
Now, let us analyze the results for the Unruh state of the field. We expect that the behavior of both $\sigma_{in}$ and $\Lambda_{nm}^{ij}$ will be different than in the case of the Hartle-Hawking state, since in contrast to the Hartle-Hawking state corresponding to the thermal bath of particles, the Unruh state of the field is perceived by a static observer as a state of outgoing radiation escaping to the asymptotic region.

We begin by analyzing the factors $\sigma_{in}$. Let us notice that in the case of the Unruh state of the field, they are closely related to the greybody factors, as they characterize the deviation of the emission spectrum from a pure black-body spectrum. However, they are not the greybody factors themselves. First of all, since they represent the dependence on the radial distance of the diagonal elements of the density matrix, simple radiation dispersion imply that they shall decay approximately as $1/r_n^2$, the approximation being better for large $r_n$. Moreover, the greybody factors are defined for each angular momentum separately, while in the definition of $\sigma_{in}$ we sum over all angular momenta. Finally, the greybody factors are defined in the asymptotic region, while we consider $\sigma_{in}$ at finite distances from the black hole. Therefore, the quantity $r_n^2\sigma_{in}$ should correspond to a `total greybody factor' for $r_n\rightarrow\infty$. The word `total' refers here to the inclusion of all angular momenta at once.

\begin{figure}
    \centering
    \includegraphics[width=0.4\textwidth]{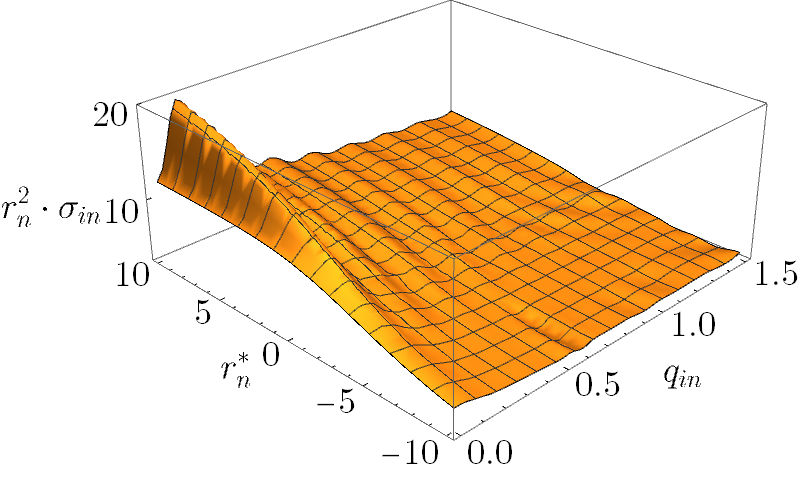}
    \caption{The value of the ``total greybody factor'' for the Unruh state, $r_n^2\sigma_{in}$, as a function of the tortoise coordinate $r_n^*$ and the energy of the excitation rescaled by appropriate blueshift factor $q_{in}$.}
    \label{fig::sigma unruh}
\end{figure}

In Fig.~\ref{fig::sigma unruh} we plot the quantity $r_n^2\sigma_{in}$ as a function of $r_n^*$ and $q_{in}$. We notice that its value increases with growing $r_n^*$, but tends to finite value both at the horizon and in the asymptotic region. On the other hand, in the $q_{in}$ direction we observe some oscillations around a decreasing value and with decreasing amplitude.

\begin{figure}
\centering
\begin{subfigure}[t]{0.4\textwidth}
         \centering
         \includegraphics[width=\textwidth]{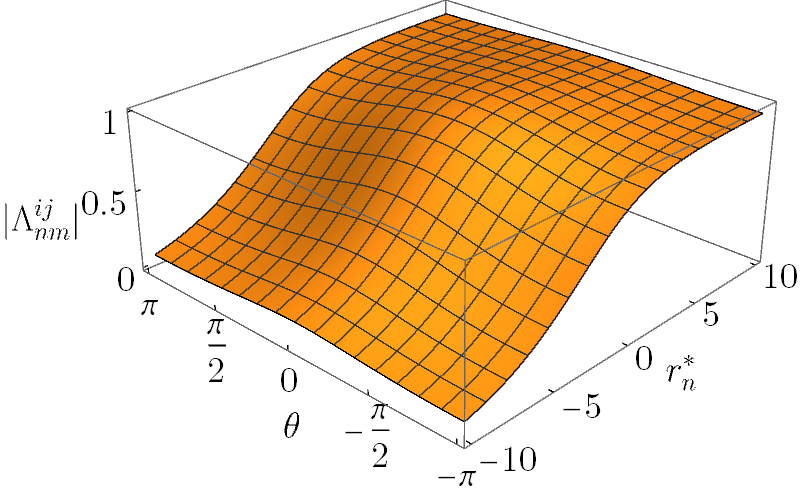}
         \caption{$q_{in}=0.01$}
         \label{fig::unruh product (a)}
\end{subfigure}
\hfill
\begin{subfigure}[t]{0.4\textwidth}
         \centering
         \includegraphics[width=\textwidth]{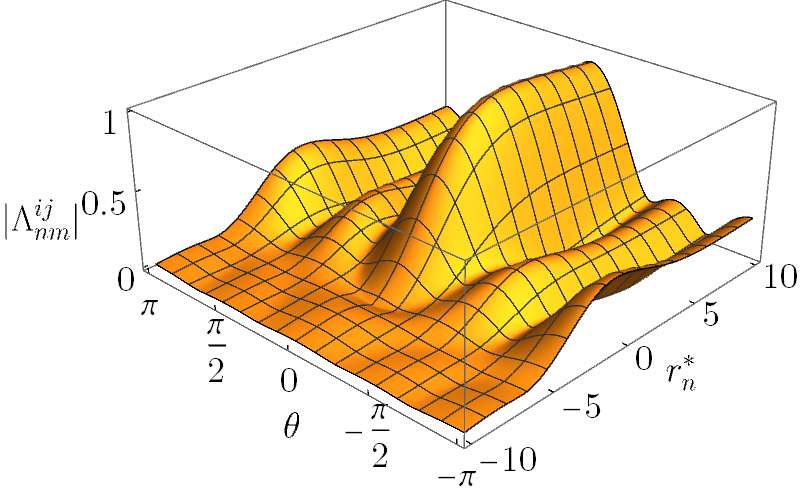}
         \caption{$q_{in}=0.5$}
         \label{fig::unruh product (b)}
\end{subfigure}
\hfill
\begin{subfigure}[t]{0.4\textwidth}
         \centering
         \includegraphics[width=\textwidth]{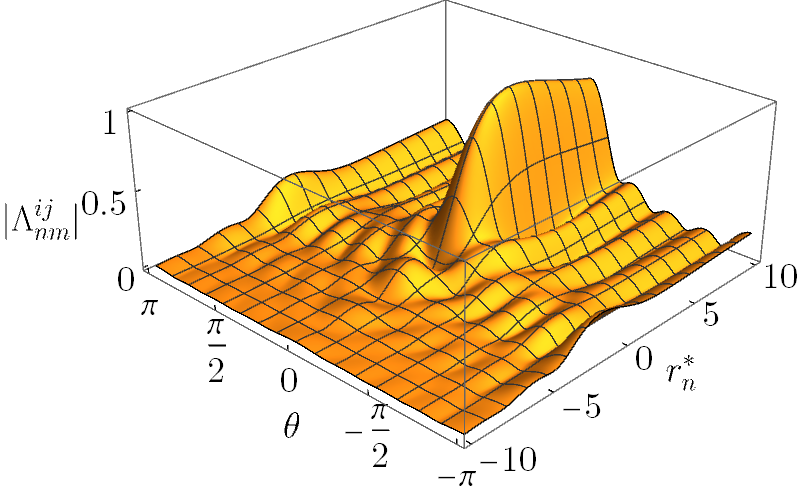}
         \caption{$q_{in}=1$}
         \label{fig::unruh product (c)}
\end{subfigure}
\hfill
\begin{subfigure}[t]{0.4\textwidth}
         \centering
         \includegraphics[width=\textwidth]{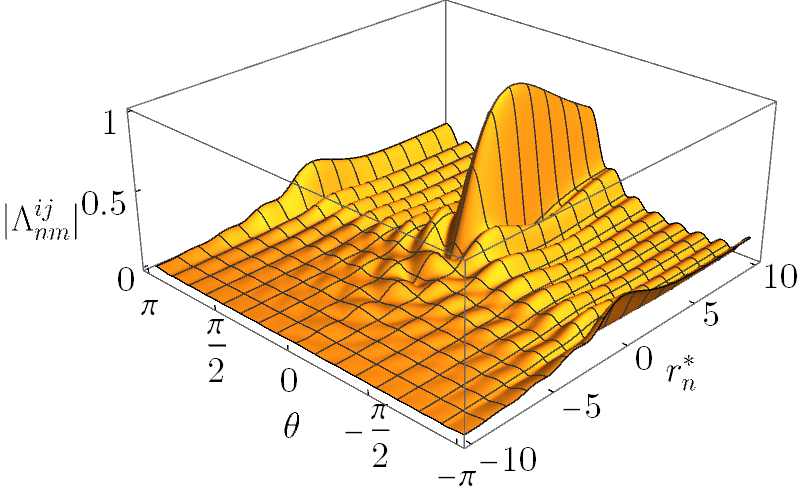}
         \caption{$q_{in}=1.5$}
         \label{fig::unruh product (d)}
\end{subfigure}
\caption{The absolute value of the normalized product $|\Lambda_{nm}^{ij}|$, for the initial state of the field being the Unruh state, as a function of the tortoise coordinate $r_n^*$ corresponding to the trajectory $\ket{n}_\text{T}$, and the angle $\theta$ formed by the trajectories $\ket{n}_\text{T}$ and $\ket{m}_\text{T}$. The radial coordinate $r_m^*$ corresponding to $\ket{m}_\text{T}$ is fixed to be $r_m^*=4M$.}
\label{fig::unruh product}
\end{figure}

Now, we consider the plots for $\Lambda_{nm}^{ij}$. We begin with Fig.~\ref{fig::unruh product}, where we fix the radial distance of the trajectory, $r_m^*=4M$, and plot $|\Lambda_{nm}^{ij}|$ as a function of the tortoise coordinate $r_n^*$ and the angle between the trajectories $\theta$. Again, the absolute value of $\Lambda_{nm}^{ij}$ reaches the maximum value of 1 only for $r_n^*=4M$ and $\theta=0$.

The functional dependence is of course even in $\theta$, but not in $r_n^*$ --- it decays to zero when we approach the horizon but tends to the finite value (different for each $\theta$) when we move away from it. For $q_{in}$ large enough, we see an oscillatory behavior in the $\theta$ direction and small oscillations in the $r_n^*$ direction. Again, quite remarkable is the presence of the relatively large coherence on the antipodes ($\theta=\pm\pi$) occurring for significantly large $q_{in}$. It is also worth noticing that the side peaks are shifted in the $r$-direction with respect to the central one.

\begin{figure*}
\centering
\begin{subfigure}[t]{0.45\textwidth}
         \centering
         \includegraphics[width=\textwidth]{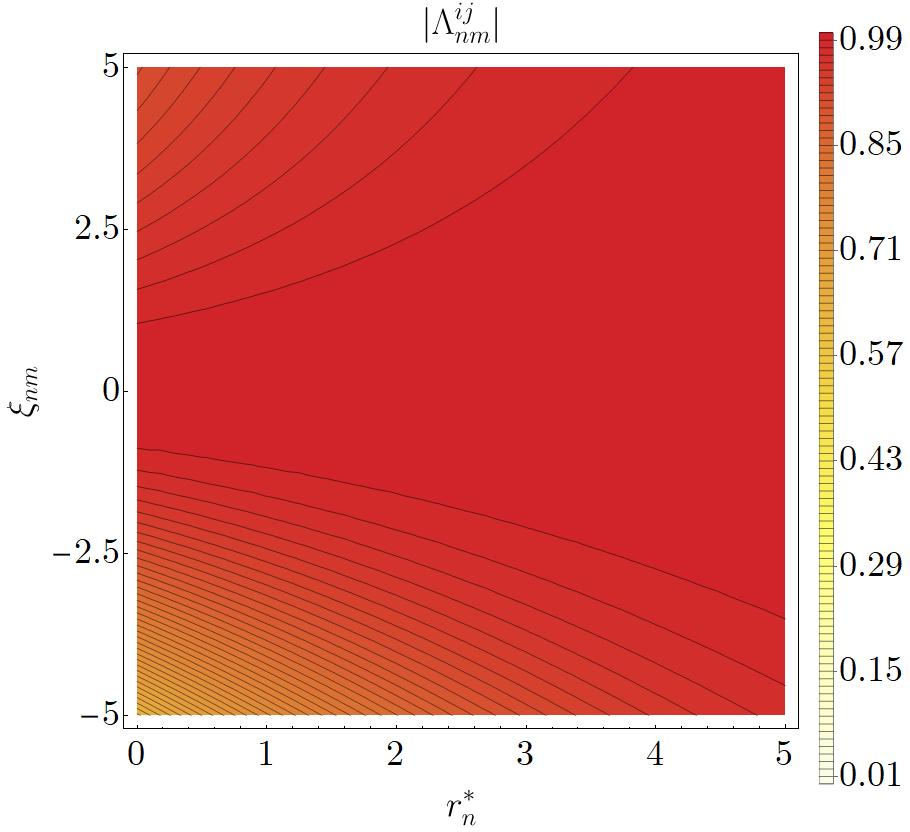}
         \caption{$q_{in}=0.01$}
         \label{fig::unruh rr product (a)}
\end{subfigure}
\hfill
\begin{subfigure}[t]{0.45\textwidth}
         \centering
         \includegraphics[width=\textwidth]{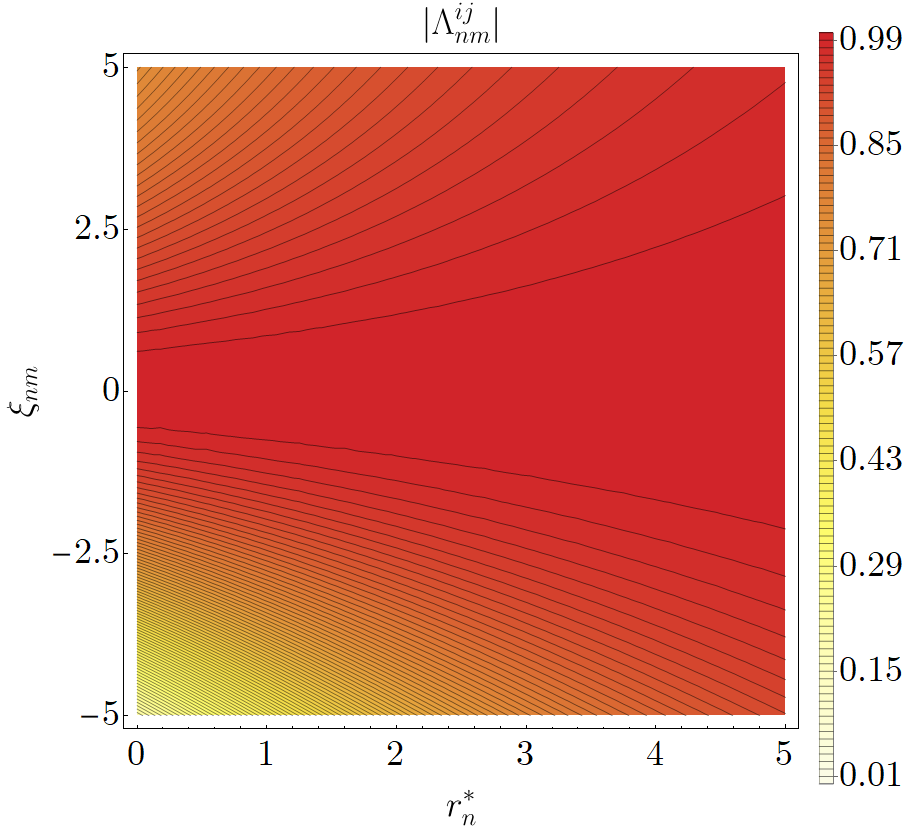}
         \caption{$q_{in}=0.5$}
         \label{fig::unruh rr product (b)}
\end{subfigure}
\begin{subfigure}[t]{0.45\textwidth}
         \centering
         \includegraphics[width=\textwidth]{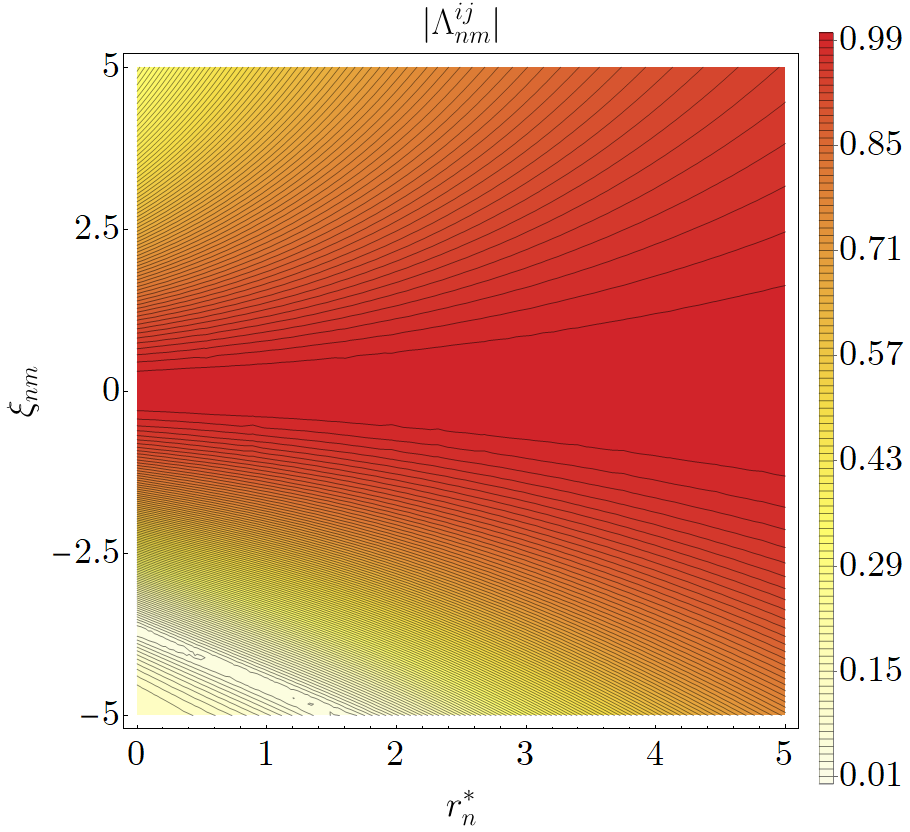}
         \caption{$q_{in}=1$}
         \label{fig::unruh rr product (c)}
\end{subfigure}
\hfill
\begin{subfigure}[t]{0.45\textwidth}
         \centering
         \includegraphics[width=\textwidth]{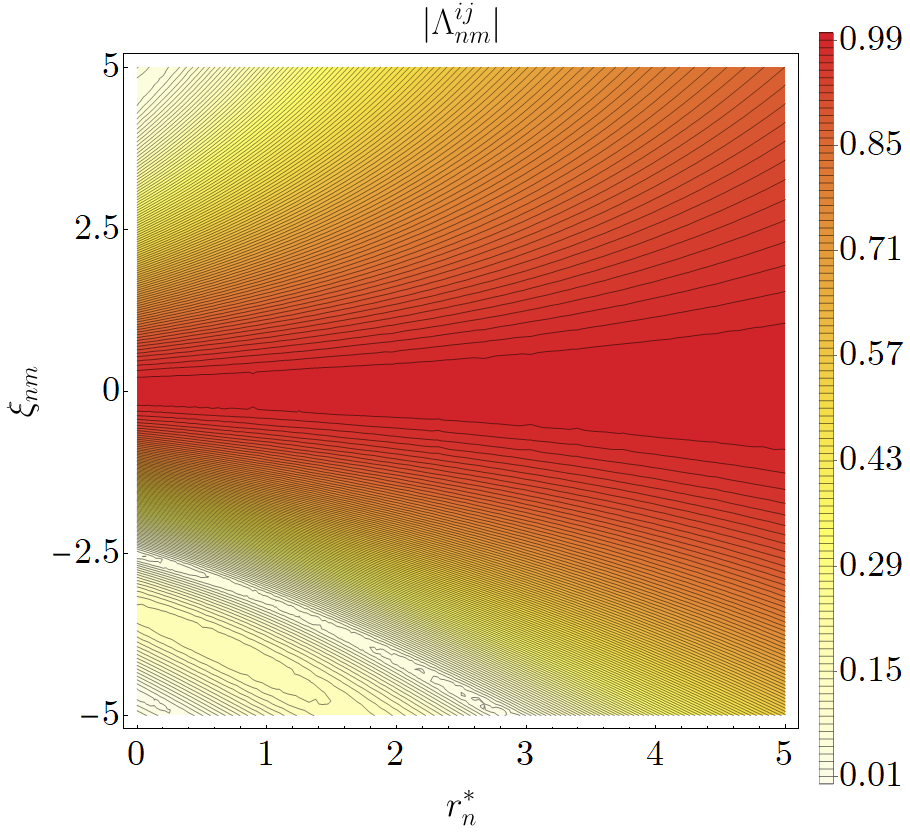}
         \caption{$q_{in}=1.5$}
         \label{fig::unruh rr product (d)}
\end{subfigure}
\caption{The absolute value of the normalized product $|\Lambda_{nm}^{ij}|$, for the initial state of the field being Unruh state, as a function of the tortoise coordinate $r_n^*$ corresponding to the trajectory $\ket{n}_\text{T}$, and the radar distance $\xi_{nm}$ between trajectories $\ket{n}_\text{T}$ and $\ket{m}_\text{T}$. The angle between the trajectories is fixed to be $\theta=0$.}
\label{fig::unruh rr product}
\end{figure*}

The plots in Fig.~\ref{fig::unruh rr product} show how the value of $\Lambda_{nm}^{ij}$ changes when we fix the angle between the trajectories $\theta=0$ and variate the radial positions of both trajectories. We make the plots of $|\Lambda_{nm}^{ij}|$ as a function of the tortoise coordinate $r_n^*$ and a radar distance $\xi_{nm}$ as defined in the previous Subsections. We see that the behavior of $\Lambda_{nm}^{ij}$ is asymmetric in $\xi_{nm}$ --- it decreases noticeably faster when we approach the black hole. For large values of $q_{in}$, we observe oscillations of the value of $|\Lambda_{nm}^{ij}|$ when we approach the horizon --- this is in agreement with the plots in Fig.~\ref{fig::unruh product}.
%

%As expected, the results are symmetric with respect to the $r_n^*=r_m^*$ axis. Based on the plots, we can see some interesting facts. First of all, for large $r_n^*$ and $r_m^*$ the absolute value of $\Lambda_{nm}^{ij}$ is close to 1 regardless of the relative distance $r_n^*-r_m^*$ between the trajectories, while for small $r_n^*$ and $r_m^*$, this value quickly drops even for small relative distances. We see that the larger the value of $q$, the larger the coordinates $r_n^*$ and $r_m^*$ for which we can still observe this drop and the sharper the drop for small values of these coordinates. On the other hand, even if we fix one of the distances to be very large, the value of $\Lambda_{nm}^{ij}$ still decreases to zero for the other distance sufficiently small. Finally, we notice that for $q\sim1$ the oscillations appear for small values of either of the distances.

Having discussed the functional dependence of $\Lambda_{nm}^{ij}$, we can deliberate on its physical meaning. Any static observer will perceive the Unruh state of the field as the radiation emitted by the black hole. According to such an observer, the detector excites when it absorbs a radiation particle. Since we assume that the detector interacts with the field for a very long time \eqref{T assumption}, the dispersion in the frequency of the excited state must be tiny --- hence the condition \eqref{condition} imposing the equality of the excitation energies as perceived by a static observer. The dependence of the scalar product $\Lambda_{nm}^{ij}$ on the angle $\theta$ for fixed $r_n^*$ and $r_m^*$ can be interpreted as the angular distribution of the emitted particle. Interestingly, for a fixed value of $r_m^*$ and $\theta$, and $r_n^*\rightarrow\infty$, the absolute value of $\Lambda_{nm}^{ij}$ does not vanish, but tends to a finite value instead. Thus, it seems that the distance at which the detector absorbs a radiation particle does not have a huge impact on the final state of the field. Notice that this is not true if the detector absorbs the particle close to the horizon since $\Lambda_{nm}^{ij}$ vanishes then. Moreover, the importance of the position at which the particle is absorbed for the final state of the field grows with the increasing value of $q_{in}$.

Finally, let us notice an interesting detail about the plots in Fig.~\ref{fig::unruh product} and Fig.~\ref{fig::hawking product} --- the number of peaks along the angular coordinate is the same in both Unruh and Hartle-Hawking state --- the plots from Fig.~\ref{fig::unruh product} resemble in appearance the plots from Fig.~\ref{fig::hawking product}, but stretched in the radial direction. It looks like the particles with the same distribution around the black hole are either moving away from it (in the case of the Unruh state) or swaying around some fixed distance (in the case of the Hartle-Hawking state). This suggests that the angular distribution of a scalar particle does not depend on the state of the field, but rather is a property of spacetime.
%--------------------------------------------
\section{Conclusions}\label{sec: conclusions}
%--------------------------------------------
We have studied the excitation of a particle detector following a quantum superposition of static trajectories in the vicinity of a Schwarzschild black hole due to Hawking radiation. We have found that the interaction with the field does not necessarily lead to total decoherence in the final state of the detector, and we have analyzed the dependence of the coherences left in the detector's state on the trajectories followed by the detector, the energy of its excitation, and the vacuum state of the field. Based on the analysis of the coherences in the final state of the detector, we have drawn conclusions regarding the properties of scalar particles in the vicinity of a static black hole.

The considerations and results in this work are in many aspects analogous to those in~\cite{Barbado}, where superpositions of uniformly accelerating trajectories were analyzed, but obtained in the significantly different scenario of the Hawking radiation in a Schwarzschild black hole. We also notice that coherences in~\cite{Barbado} appeared only when considering highly fine-tuned hyperbolic trajectories sharing almost exactly the same Rindler wedge. In contrast, in the present work coherences appear for the much more natural setting of superpositions of static trajectories outside a black hole. Therefore, while in both cases the superposition requires the same order of fine-tuning, in the second scenario such fine-tuning appears much more reasonable to achieve.

Several possibilities for further research in the direction set by this work naturally arise. First of all, it would be relevant to interpret our results within the framework of the quantum reference frames \cite{Guerin, Giacomini}. In this framework, one associates a reference frame to a physical system, which can have quantum features. For example, if we consider a reference frame in which a classical observer perceives a quantum system to be in a superposition, we can associate another reference frame with the quantum system, in which the ``classical'' observer will be in a superposition, and the quantum system will follow a well-defined trajectory. From that perspective, an observer following a quantum superposition of trajectories in Schwarzschild spacetime would consider himself to be static near a black hole in a superposition of locations.

Another idea would be to consider other types of trajectories beyond the static ones considered in this work. Based on the paper \cite{Hodgkinson1}, we think that it should be relatively straightforward to treat circular trajectories, as it requires only minor modifications in our calculations. Yet another thing to do could be to consider detectors following superpositions of trajectories in other spacetimes. Based on \cite{Hodgkinson2} one can consider the analogous problem on the BTZ black hole. Considering other spacetimes like Kerr or Reissner-Nordström would probably require much more effort, but should also be achievable. The scalar field treatment in these spacetimes was given in \cite{Ford} for Kerr, and in \cite{Castineiras} for Reissner-Nordström metric.
%--------------------------------------------
\begin{acknowledgments}
The authors want to thank \v{C}aslav Brukner and Andrzej Dragan for useful comments and discussions during the elaboration of this article. L.C.B.\ acknowledges the support from the research platform TURIS, from the \"{O}AW through the project ``Quantum Reference Frames for Quantum Fields'' (ref.~IF\textunderscore 2019\textunderscore 59\textunderscore QRFQF), and from the European Commission via Testing the Large-Scale Limit of Quantum Mechanics (TEQ) (No. 766900) project, also supported by the Austrian-Serbian bilateral scientific cooperation no. 451-03-02141/2017-09/02, and by the Austrian Science Fund (FWF) through the SFB project BeyondC (sub-project~F7103) and a grant from the Foundational Questions Institute (FQXi) Fund. This publication was made possible through the support of the ID61466 grant from the John Templeton Foundation, as part of the The Quantum Information Structure of Spacetime (QISS) Project (qiss.fr). The opinions expressed in this publication are those of the author(s)and do not necessarily reflect the views of the John Templeton Foundation.

\end{acknowledgments}
%--------------------------------------------

\onecolumngrid
\newpage
\appendix

%--------------------------------------------
\section{Computation of the scalar products of the states of the field}\label{Appendix A}
%--------------------------------------------

In this Appendix, we provide a detailed computation of the scalar product $\braket{\omega_i,n}{\omega_j,m}_\text{F}$. Let us explicitly write the state $\ket{\omega_i,n}_\text{F}$ by including the form of the interaction Hamiltonian \eqref{interaction hamiltonian} in the formula for the state of the field \eqref{field state}
\begin{equation}
\begin{split}
    \ket{\omega_i,n}_\text{F}=&(\varepsilon\zeta_i)^{-1}\bra{\omega_i}_\text{D}\bra{n}_\text{T}\hat{S}_I\ket{0}_\text{D}\ket{\Omega}_\text{F}\ket{n}_\text{T}\\
    =&(\varepsilon\zeta_i)^{-1}\bra{\omega_i}_\text{D}\bra{n}_\text{T}\varepsilon\int_{-\infty}^\infty \mathrm{d}\tau\chi(\tau)\hat{m}(\tau)\hat{\phi}(\hat{x}(\tau))\ket{0}_\text{D}\ket{\Omega}_\text{F}\ket{n}_\text{T}\\
    =&\int_{-\infty}^\infty\mathrm{d}\tau\chi(\tau)\zeta_i^{-1}\bra{\omega_i}\hat{m}(\tau)\ket{0}_\text{D}\hat{\phi}(x)\ket{\Omega}_F\\
    =&\int_{-\infty}^\infty\mathrm{d}\tau\chi(\tau)\mathrm{e}^{i\omega_i\tau}\hat{\phi}(x)\ket{\Omega}_F.
\end{split}
\end{equation}
Equipped with this formula, we can write the product $\braket{\omega_i,n}{\omega_j,m}_\text{F}$ in the compact form
\begin{equation}\label{product0}
    \braket{\omega_i,n}{\omega_j,m}_\text{F}=\int_{-\infty}^\infty\mathrm{d}\tau\int_{-\infty}^\infty\mathrm{d}\tau'\chi(\tau)\chi(\tau')\mathrm{e}^{-\mathrm{i}(\omega_i\tau-\omega_j\tau')}W(x_n,x_m)
\end{equation}
with
\begin{equation}\label{trajectories}
\begin{split}
    &x_n=\left(\frac{\tau}{\sqrt{1-\frac{2M}{r_n}}},r_n,0,0\right)\equiv\left(\alpha_n\tau,r_n,0,0\right),\\
    &x_m=\left(\frac{\tau}{\sqrt{1-\frac{2M}{r_m}}},r_m,\theta,0\right)\equiv(\alpha_m\tau',r_m,\theta,0),
\end{split}
\end{equation}
being the trajectories corresponding to $\ket{n}_\text{T}$ and $\ket{m}_\text{T}$, respectively. Without loss of generality, we have fixed to zero the azimuthal angle of both trajectories and the polar angle of the first trajectory --- we can do that because of the spherical symmetry of the metric \eqref{schwarzschild metric}. The quantity $W(x_n,x_m)$ is the Wightman function, defined by
\begin{equation}
    W(x_n,x_m)\equiv\bra{\Omega}\hat{\phi}(x_n)\hat{\phi}(x_m)\ket{\Omega}_\text{F};
\end{equation}
and it clearly depends on the initial state of the field $\ket{\Omega}_\text{F}$. The Wightman function has already been computed in the literature for $\ket{\Omega}_\text{F}$ being either the Unruh, or the Hartle-Hawking state \cite{Hodgkinson1}. It is given by
\begin{equation}\label{unruh}
\begin{split}
W_U\left(x_n,x_m\right)=& \sum_{\ell=0}^{\infty} \sum_{\mu=-\ell}^{+\ell} \int_0^{\infty} \mathrm{d} \omega\left[\frac{\mathrm{e}^{4 \pi M \omega-\mathrm{i} \omega \Delta t} Y_{\ell \mu}(0, 0) Y_{\ell \mu}^*\left(\theta, 0\right) \Phi_{\omega \ell}^{\mathrm{up}}(r_n) \Phi_{\omega \ell}^{\mathrm{up} *}\left(r_m\right)}{8 \pi \omega \sinh (4 \pi M \omega)}\right.\\
&+\frac{\mathrm{e}^{-4 \pi M \omega+\mathrm{i} \omega \Delta t} Y_{\ell \mu}^*\left(0, 0\right) Y_{\ell \mu}\left(\theta,0\right) \Phi_{\omega \ell}^{\mathrm{up} *}(r_n) \Phi_{\omega \ell}^{\mathrm{up}}\left(r_m\right)}{8 \pi \omega \sinh (4 \pi M \omega)}\\
&\left.+\frac{\mathrm{e}^{-\mathrm{i} \omega \Delta t} Y_{\ell \mu}(0, 0) Y_{\ell \mu}^*\left(\theta, 0\right) \Phi_{\omega \ell}^{\mathrm{in}}(r_n) \Phi_{\omega \ell}^{\mathrm{in} *}\left(r_m\right)}{4 \pi \omega}\right]
\end{split}
\end{equation}
if the initial state of the field is the Unruh state, and by
\begin{equation}\label{hartle_hawking}
\begin{split}
W_H\left(x_n,x_m\right)
=& \sum_{\ell=0}^{\infty} \sum_{\mu=-\ell}^{+\ell} \int_0^{\infty} \mathrm{d} \omega \frac{1}{8 \pi \omega \sinh (4 \pi M \omega)} \\
& \times\left[\mathrm{e}^{4 \pi M \omega-\mathrm{i} \omega \Delta t} Y_{\ell \mu}(0, 0) Y_{\ell \mu}^*\left(\theta, 0\right)\left(\Phi_{\omega \ell}^{\mathrm{up}}(r_n) \Phi_{\omega \ell}^{\mathrm{up} *}(r_m)+\Phi_{\omega \ell}^{\mathrm{in}}(r_n) \Phi_{\omega \ell}^{\mathrm{in} *}(r_m)\right)\right.\\
&\left.+\mathrm{e}^{-4 \pi M \omega+\mathrm{i} \omega \Delta t} Y_{\ell \mu}^*\left(0, 0\right) Y_{\ell \mu}(\theta, 0)\left(\Phi_{\omega \ell}^{\mathrm{up} *}(r_n) \Phi_{\omega \ell}^{\mathrm{up}}(r_m)+\Phi_{\omega \ell}^{\mathrm{in} *}(r_n) \Phi_{\omega \ell}^{\mathrm{in}}(r_m)\right)\right]
\end{split}
\end{equation}
if the initial state of the field is the Hartle-Hawking state. Here $\Delta t=\alpha_n\tau-\alpha_m\tau'$, the functions $Y_{\ell \mu}$ are the spherical harmonics, and $\Phi_{\omega \ell}^{\mathrm{up}}$ and $\Phi_{\omega \ell}^{\mathrm{in}}$ stand for the normalized solutions of the differential equation
\begin{equation}\label{modes}
\phi_{\omega \ell}^{\prime \prime}(r)+\frac{2(r-M)}{r(r-2 M)} \phi_{\omega \ell}^{\prime}(r)
+\left(\frac{\omega^2 r^2}{(r-2 M)^2}-\frac{\ell(\ell+1)}{r(r-2 M)}\right) \phi_{\omega \ell}(r)=0
\end{equation}
with specified asymptotic behavior at infinity and at the horizon, respectively. The equation \eqref{modes} cannot be solved analytically. The numerical procedure for computing $\Phi_{\omega \ell}^{\mathrm{up}}$ and $\Phi_{\omega \ell}^{\mathrm{in}}$ is described in detail in \cite{Hodgkinson1}.

\subsection{Hatrle-Hawking state}\label{Appendix A - Hawking}
%--------------------------------------------
Inserting \eqref{hartle_hawking} to \eqref{product0} and making use of the fact that $Y_{\ell \mu}(0,0)=\sqrt{\frac{2\ell+1}{4\pi}}\delta_{\mu 0}$ (see \cite{nist}), we get
\begin{equation}\label{product u}
\begin{split}
  \braket{\omega_i,n}{\omega_j,m}_\text{F}=\int_{-\infty}^\infty\mathrm{d}\tau\int_{-\infty}^\infty&\mathrm{d}\tau'\chi(\tau)\chi(\tau')\mathrm{e}^{-\mathrm{i}(\omega_i\tau-\omega_j\tau')}\sum_{\ell=0}^{\infty}  \int_0^{\infty} \mathrm{d} \omega\frac{1}{8 \pi \omega \sinh (4 \pi M \omega)}\times\\
  \times\sqrt{\frac{2\ell+1}{4\pi}}&\left[\mathrm{e}^{4 \pi M \omega-\mathrm{i} \omega \Delta t} Y_{\ell \mu}^*\left(\theta, 0\right)\left(\Phi_{\omega \ell}^{\mathrm{up}}(r_n) \Phi_{\omega \ell}^{\mathrm{up} *}(r_m)+\Phi_{\omega \ell}^{\mathrm{in}}(r_n) \Phi_{\omega \ell}^{\mathrm{in} *}(r_m)\right)\right.\\
&\left.+\mathrm{e}^{-4 \pi M \omega+\mathrm{i} \omega \Delta t} Y_{\ell \mu}(\theta, 0)\left(\Phi_{\omega \ell}^{\mathrm{up} *}(r_n) \Phi_{\omega \ell}^{\mathrm{up}}(r_m)+\Phi_{\omega \ell}^{\mathrm{in} *}(r_n) \Phi_{\omega \ell}^{\mathrm{in}}(r_m)\right)\right].
\end{split}
\end{equation}
This can be further simplified, if we recall that $Y_{\ell 0}(\theta,0)=\sqrt{\frac{(2\ell+1)}{4\pi}}P_\ell(\cos\theta)$, where $P_\ell(x)$ is the Legendre polynomial. With this formula at hand, we rewrite \eqref{product u} as
\begin{equation}\label{product u0}
\begin{split}
  \braket{\omega_i,n}{\omega_j,m}_\text{F}=\int_{-\infty}^\infty\mathrm{d}\tau\int_{-\infty}^\infty&\mathrm{d}\tau'\chi(\tau)\chi(\tau')\mathrm{e}^{-\mathrm{i}(\omega_i\tau-\omega_j\tau')}\sum_{\ell=0}^{\infty}  \int_0^{\infty} \mathrm{d} \omega\frac{P_\ell(\cos\theta)}{8 \pi \omega \sinh (4 \pi M \omega)}\times\\
  \times\frac{2\ell+1}{4\pi}&\left[\mathrm{e}^{4 \pi M \omega-\mathrm{i} \omega \Delta t} \left(\Phi_{\omega \ell}^{\mathrm{up}}(r_n) \Phi_{\omega \ell}^{\mathrm{up} *}(r_m)+\Phi_{\omega \ell}^{\mathrm{in}}(r_n) \Phi_{\omega \ell}^{\mathrm{in} *}(r_m)\right)\right.\\
&\left.+\mathrm{e}^{-4 \pi M \omega+\mathrm{i} \omega \Delta t} \left(\Phi_{\omega \ell}^{\mathrm{up} *}(r_n) \Phi_{\omega \ell}^{\mathrm{up}}(r_m)+\Phi_{\omega \ell}^{\mathrm{in} *}(r_n) \Phi_{\omega \ell}^{\mathrm{in}}(r_m)\right)\right].
\end{split}
\end{equation}
Let us notice that the only $\tau$-, and $\tau'$-dependence in the big bracket comes from the factors $\mathrm{e}^{\pm\mathrm{i}\omega\Delta t}$, therefore the integrals over $\tau$ and $\tau'$ can be performed. Indeed, we can define
\begin{equation}
    \int_{-\infty}^\infty \mathrm{d}\tau\chi(\tau)\mathrm{e}^{i\Omega\tau}=(8\pi)^{\frac{1}{4}}T\mathrm{e}^{-\Omega^2 T^2}=:\tilde{\chi}(\Omega),
\end{equation}
where $\tilde{\chi}(\Omega)$ denotes the Fourier transform of $\chi(\tau)$. Hence,
\begin{equation}\label{tau_integral1}
\begin{split}
    &\int_{-\infty}^\infty\mathrm{d}\tau\int_{-\infty}^\infty\mathrm{d}\tau'\chi(\tau)\chi(\tau')\mathrm{e}^{\mathrm{i}(\omega_j\tau'-\omega_i\tau)}\mathrm{e}^{\pm \mathrm{i}\omega\Delta t}\\
    =&\int_{-\infty}^\infty\mathrm{d}\tau\int_{-\infty}^\infty\mathrm{d}\tau'\chi(\tau)\chi(\tau')\mathrm{e}^{\mathrm{i}(\omega_j\tau'-\omega_i\tau)}\mathrm{e}^{\pm \mathrm{i}\omega(\alpha_n\tau-\alpha_m\tau')}\\
    =&\int_{-\infty}^\infty\mathrm{d}\tau\chi(\tau)\mathrm{e}^{-\mathrm{i}(\omega_i\mp \omega\alpha_n)\tau}\int_{-\infty}^\infty\mathrm{d}\tau'\chi(\tau')\mathrm{e}^{\mathrm{i}(\omega_j\mp\alpha_m\omega)\tau'}\\
    =&\tilde{\chi}(-\omega_i\pm\omega\alpha_n)\tilde{\chi}(\omega_j\mp\omega\alpha_m)=\tilde{\chi}(\omega_i\mp\omega\alpha_n)\tilde{\chi}(\omega_j\mp\omega\alpha_m),
\end{split}
\end{equation}
where in the first step we used the definition of $\Delta t$, and in the last one we made use of the fact that $\tilde{\chi}(-\Omega)=\tilde{\chi}(\Omega)$.

Now, we can make use of the adiabaticity condition \eqref{T assumption}. Since we assume that the switching time $T$ is very large compared to other time scales appearing in the problem ($T\gg 1/\omega_1\ge1/\omega_i$), the function $\tilde{\chi}(\Omega)$ is very sharp. Therefore, the product $\tilde{\chi}(\Omega_1)\tilde{\chi}(\Omega_2)$ is non-negligible only for $\Omega_1\approx\Omega_2$. In our case, this leads to the condition
\begin{equation}\label{omega_condition}
\frac{\omega_i}{\alpha_n}\approx\frac{\omega_j}{\alpha_m},
\end{equation}
meaning that the excitation energies divided by the blueshift factors must be the same on both trajectories. Assuming that the above equality holds to order $\varepsilon$ and denoting $q_{in}:=\frac{\omega_i}{\alpha_n}$, we rewrite \eqref{tau_integral1} in the form
\begin{equation}\label{tau_integral2}
    \tilde{\chi}(\omega_i\mp\omega\alpha_n)\tilde{\chi}(\omega_j\mp\omega\alpha_m)=\tilde{\chi}(\alpha_n(q_{in}\mp\omega))\tilde{\chi}(\alpha_m(q_{in}\mp\omega))=\sqrt{8\pi}T^2\mathrm{e}^{-(\omega\mp q_{in})^2 (\alpha_n^2+\alpha_m^2)T^2}.
\end{equation}
Introducing this result in \eqref{product u0} we get
\begin{equation}\label{product u1}
\begin{split}
  \braket{\omega_i,n}{\omega_j,m}_\text{F}=T^2&\sum_{\ell=0}^{\infty} \frac{2\ell+1}{4\pi}P_\ell\left(\cos\theta\right) \int_0^{\infty} \mathrm{d} \omega \frac{1}{2 \sqrt{2\pi} \omega \sinh (4 \pi M \omega)} \\
\times&\left[\mathrm{e}^{-(\omega + q_{in})^2 (\alpha_n^2+\alpha_m^2)T^2+4 \pi M \omega}  \left(\Phi_{\omega \ell}^{\mathrm{up}}(r_n) \Phi_{\omega \ell}^{\mathrm{up} *}(r_m)+\Phi_{\omega \ell}^{\mathrm{in}}(r_n) \Phi_{\omega \ell}^{\mathrm{in} *}(r_m)\right)\right.\\
&\left.+\mathrm{e}^{-(\omega - q_{in})^2 (\alpha_n^2+\alpha_m^2)T^2 - 4 \pi M \omega} \left(\Phi_{\omega \ell}^{\mathrm{up} *}(r_n) \Phi_{\omega \ell}^{\mathrm{up}}(r_m)+\Phi_{\omega \ell}^{\mathrm{in} *}(r_n) \Phi_{\omega \ell}^{\mathrm{in}}(r_m)\right)\right].
\end{split}
\end{equation}

Once again, we can use the fact that $T$ is large to perform the integral over $\omega$. Laplace's method tells us that for a positive function $h(\omega)$, a large number $N\to\infty$, and a twice-differentiable function $g(\omega)$ with a unique global maximum at $\omega_0\in(a,b)$, the integral can be approximated as
\begin{equation}\label{laplace's method}
\int_a^b h(\omega) e^{N g(\omega)} d \omega \approx \sqrt{\frac{2 \pi}{N\left|g^{\prime \prime}\left(\omega_0\right)\right|}} h\left(\omega_0\right) e^{N g\left(\omega_0\right)}.
\end{equation}
In the integrals in \eqref{product u1} $N=(\alpha_n^2+\alpha_m^2)T^2$, and $g(\omega)=-(\omega\pm q_{in})^2$. Notice that $g(\omega)$ has a unique global maximum at $\omega_0 = \mp q_{in}$, which is in the range of integration $(a,b)=(0,\infty)$ only for $g(\omega)=-(\omega-q_{in})^2$ (recall that $q_{in}=\omega_i/\alpha_n > 0$). Thus, the terms with $g(\omega)=-(\omega+q_{in})^2$ vanish, and we are left with
\begin{equation}\label{product u2}
\begin{split}
  \braket{\omega_i,n}{\omega_j,m}_\text{F}=&\frac{T}{q_{in} \sqrt{2(\alpha_n^2+\alpha_m^2)}}\frac{1}{\mathrm{e}^{q_{in}/T_\text{H}}-1}\sum_{\ell=0}^{\infty} \frac{2\ell+1}{4\pi} P_\ell\left(\cos\theta\right)\times\\ 
  &\times\left(\Phi_{q_{in} \ell}^{\mathrm{up} *}(r_n) \Phi_{q_{in} \ell}^{\mathrm{up}}(r_m)+\Phi_{q_{in} \ell}^{\mathrm{in} *}(r_n) \Phi_{q_{in} \ell}^{\mathrm{in}}(r_m)\right),
\end{split}
\end{equation}
where $T_\text{H}$ is the Hawking temperature \eqref{Hawking temperature}.

%\luis{[Doesn't the thermal factor appear because: First, you compute the diagonal terms, for which you find the thermal factor; and second, for the non-diagonal terms, you multiply and divide by the square root of the product of two diagonal terms (see our procedure), obtaining the thermal factor on one side (because you just multiply by it), and a normalized inner product (our~$\Lambda$, because you divided). If this is so, then it's better to write it like that.]}

In order to compute the diagonal terms ($n=m$, and $\omega_i=\omega_j$), we set $\theta=0$, $r=r_m$, and $\alpha_n=\alpha_m$. After introducing this in \eqref{product u2}, we get
\begin{equation}
    \braket{\omega_i,n}{\omega_i,n}_\text{F}=\frac{T}{8\pi q_{in}\alpha_n}\frac{1}{\mathrm{e}^{q_{in}/T_\text{H}}-1}\sum_{\ell=0}^{\infty} (2\ell+1) \left(\left|\Phi_{q_{in} \ell}^{\mathrm{up}}(r_n)\right|^2+\left|\Phi_{q_{in} \ell}^{\mathrm{in}}(r_n)\right|^2\right)\equiv\frac{T}{2\pi}\frac{\sigma_{in}\omega_i}{\mathrm{e}^{q_{in}/T_\text{H}}-1},
\end{equation}
where we defined
\begin{equation}\label{sigma hartle-hawking}
    \sigma_{in}:=\frac{2\pi}{\omega_iT}\left(\mathrm{e}^{q_{in}/T_\text{H}}-1\right)\braket{\omega_i,n}{\omega_i,n}_\text{F}=\frac{1}{4 \omega_i^2}\sum_{\ell=0}^{\infty} (2\ell+1)\left(\left|\Phi_{q_{in} \ell}^{\mathrm{up}}(r_n)\right|^2+\left|\Phi_{q_{in} \ell}^{\mathrm{in}}(r_n)\right|^2\right).
\end{equation}
Finally, the inner product between normalized states of the field is given by
\begin{equation}
\begin{split}
    \Lambda_{nm}^{ij}:=&\frac{\braket{\omega_i,n}{\omega_j,m}_\text{F}}{\sqrt{\braket{\omega_i,n}{\omega_i,n}_\text{F}\braket{\omega_j,m}{\omega_j,m}_\text{F}}}\\
    =&\frac{\sqrt{2\frac{\alpha_n\alpha_m}{(\alpha_n^2+\alpha_m^2)}}\sum_{\ell=0}^{\infty} (2\ell+1) P_\ell\left(\cos\theta\right) \left(\Phi_{q_{in} \ell}^{\mathrm{up} *}(r_n) \Phi_{q_{in} \ell}^{\mathrm{up}}(r_m)+\Phi_{q_{in} \ell}^{\mathrm{in} *}(r_n) \Phi_{q_{in} \ell}^{\mathrm{in}}(r_m)\right)}{\left[\sum_{\ell,\ell'=0}^{\infty} (2\ell+1)(2\ell'+1) \left(\left|\Phi_{q_{in} \ell}^{\mathrm{up}}(r_n)\right|^2+\left|\Phi_{q_{in} \ell}^{\mathrm{in}}(r_n)\right|^2\right)\left(\left|\Phi_{q_{in} \ell'}^{\mathrm{up}}(r_m)\right|^2+\left|\Phi_{q_{in} \ell'}^{\mathrm{in}}(r_m)\right|^2\right)\right]^{1/2}}.
\end{split}
\end{equation}
%

%\luis{Is it not possible to further compute the sum? This factor given by the sum is clearly the grey-body factor. We should mention that at some point.}

%--------------------------------------------
\subsection{Unruh state}\label{Appendix A - Unruh}
%--------------------------------------------

%\luis{[Clearly the previous comments also apply to this section.]}

In the case of the Unruh state, the procedure is analogous. Inserting \eqref{unruh} to \eqref{product0}, and using $Y_{\ell \mu}(0,0)=\sqrt{\frac{2\ell+1}{4\pi}}\delta_{\mu 0}$ and $Y_{\ell 0}(\theta,0)=\sqrt{\frac{(2\ell+1)}{4\pi}}P_\ell(\cos\theta)$, we get
\begin{equation}\label{product h0}
\begin{split}
  \braket{\omega_i,n}{\omega_j,m}_\text{F}=&\int_{-\infty}^\infty\mathrm{d}\tau\int_{-\infty}^\infty\mathrm{d}\tau'\chi(\tau)\chi(\tau')\mathrm{e}^{-\mathrm{i}(\omega_i\tau-\omega_j\tau')}\sum_{\ell=0}^{\infty}  \int_0^{\infty} \mathrm{d} \omega P_\ell\left(\cos\theta\right)\frac{2\ell+1}{4\pi}\times\\
  \times&\left[\frac{\mathrm{e}^{4 \pi M \omega-\mathrm{i} \omega \Delta t} \Phi_{\omega \ell}^{\mathrm{up}}(r_n) \Phi_{\omega \ell}^{\mathrm{up} *}\left(r_m\right)}{8 \pi \omega \sinh (4 \pi M \omega)}+\frac{\mathrm{e}^{-4 \pi M \omega+\mathrm{i} \omega \Delta t} \Phi_{\omega \ell}^{\mathrm{up} *}(r_n) \Phi_{\omega \ell}^{\mathrm{up}}\left(r_m\right)}{8 \pi \omega \sinh (4 \pi M \omega)}\right.\\
  &\left.+\frac{\mathrm{e}^{-\mathrm{i} \omega \Delta t} \Phi_{\omega \ell}^{\mathrm{in}}(r_n) \Phi_{\omega \ell}^{\mathrm{in} *}\left(r_m\right)}{4 \pi \omega}\right].
\end{split}
\end{equation}
We perform the integrals over $\tau$ and $\tau'$ like in the case of the Hartle-Hawking state, and obtain
\begin{equation}\label{product h1}
\begin{split}
  \braket{\omega_i,n}{\omega_j,m}_\text{F}=&T^2\sum_{\ell=0}^{\infty} \frac{2\ell+1}{4\pi}P_\ell\left(\cos\theta\right) \int_0^{\infty} \mathrm{d} \omega \left[\frac{\mathrm{e}^{-(\omega + q)^2 (\alpha_n^2+\alpha_m^2)T^2+4 \pi M \omega}  \Phi_{\omega \ell}^{\mathrm{up}}(r_n) \Phi_{\omega \ell}^{\mathrm{up} *}\left(r_m\right)}{2 \sqrt{2\pi} \omega \sinh (4 \pi M \omega)}\right.\\
&+\left.\frac{\mathrm{e}^{-(\omega - q)^2 (\alpha_n^2+\alpha_m^2)T^2 - 4 \pi M \omega} \Phi_{\omega \ell}^{\mathrm{up} *}(r_n) \Phi_{\omega \ell}^{\mathrm{up}}\left(r_m\right)}{2 \sqrt{2\pi} \omega \sinh (4 \pi M \omega)}+\frac{\mathrm{e}^{-(\omega + q)^2 (\alpha_n^2+\alpha_m^2)T^2} \Phi_{\omega \ell}^{\mathrm{in}}(r_n) \Phi_{\omega \ell}^{\mathrm{in} *}\left(r_m\right)}{\sqrt{2\pi} \omega}\right].
\end{split}
\end{equation}
Again, we use Laplace's method to perform the integral over $\omega$. This gives us
\begin{equation}\label{product h2}
  \braket{\omega_i,n}{\omega_j,m}_\text{F}=\frac{T}{q \sqrt{2(\alpha_n^2+\alpha_m^2)}}\frac{1}{\mathrm{e}^{q/T_\text{H}}-1}\sum_{\ell=0}^{\infty} \frac{2\ell+1}{4\pi} P_\ell\left(\cos\theta\right) \Phi_{q \ell}^{\mathrm{up} *}(r_n) \Phi_{q \ell}^{\mathrm{up}}\left(r_m\right).
\end{equation}
For the diagonal terms we get
\begin{equation}
    \braket{\omega_i,n}{\omega_i,n}_\text{F}=\frac{T}{8\pi q\alpha_n}\frac{1}{\mathrm{e}^{q/T_\text{H}}-1}\sum_{\ell=0}^{\infty} (2\ell+1) \left|\Phi_{q \ell}^{\mathrm{up}}(r_n)\right|^2\equiv\frac{T}{2\pi}\frac{\sigma_{in}\omega_i}{\mathrm{e}^{q/T_\text{H}}-1},
\end{equation}
with
\begin{equation}\label{sigma unruh}
    \sigma_{in}=\frac{1}{4 \omega_i^2}\sum_{\ell=0}^{\infty} (2\ell+1) \left|\Phi_{q \ell}^{\mathrm{up}}(r_n)\right|^2.
\end{equation}
Now, the inner product between the normalized states of the field reads
\begin{equation}
    \Lambda_{nm}^{ij}=\sqrt{\frac{2\alpha_n\alpha_m}{(\alpha_n^2+\alpha_m^2)}}\frac{\sum_{\ell=0}^{\infty} (2\ell+1)P_\ell\left(\cos\theta\right) \Phi_{q \ell}^{\mathrm{up} *}(r_n) \Phi_{q \ell}^{\mathrm{up}}\left(r_m\right)}{\left(\sum_{\ell,\ell'=0}^{\infty} (2\ell+1)(2\ell'+1) \left|\Phi_{q \ell}^{\mathrm{up}}(r_n)\Phi_{q \ell'}^{\mathrm{up}}(r_m)\right|^2\right)^{1/2}}.
\end{equation}
%--------------------------------------------
\subsection{Computing the up-modes and in-modes}
%--------------------------------------------
In this section we briefly describe the procedure for computing the modes $\Phi_{\omega \ell}^{\mathrm{up}}$ and $\Phi_{\omega \ell}^{\mathrm{in}}$ given in detail in \cite{Hodgkinson1}. These are the solutions of Eq.~\eqref{modes} with specified behavior at the boundary
\begin{equation}
    \begin{split}
        &\Phi_{\omega \ell}^{\mathrm{up}}(r_n)\sim\frac{\mathrm{e}^{\mathrm{i}\omega r_n^*}}{r}\qquad \text{for }r\to\infty,\\
        &\Phi_{\omega \ell}^{\mathrm{in}}(r_n)\sim\frac{\mathrm{e}^{-\mathrm{i}\omega r_n^*}}{r}\qquad \text{for }r\to2M,
    \end{split}
\end{equation}
where $r_n^*$ is a tortoise coordinate
\begin{equation}
    r_n^*=r+2M\log(r/2M-1).
\end{equation}
For the moment, let us work with arbitrary (not normalized) modes $\phi_{\omega \ell}^{\mathrm{up}}$, and $\phi_{\omega \ell}^{\mathrm{in}}$, and impose the normalization at the end. To obtain their value, we solve Eq.~\eqref{modes} numerically, setting appropriate boundary conditions for the up-modes and in-modes.

%--------------------------------------------
\subsubsection{Boundary conditions for the up-modes}
%--------------------------------------------
For the $\phi_{\omega \ell}^{\mathrm{up}}$ modes, we fix the boundary conditions at infinity
\begin{equation}
    \begin{split}
        &\phi_{\omega \ell}^{\mathrm{up}}(r_\text{inf})=\left.\frac{\mathrm{e}^{\mathrm{i}\omega r_n^*}}{r}\mathrm{e}^{v(r_n)}\right|_{r=r_\text{inf}},\\
        &\phi_{\omega \ell}^{\mathrm{up}\prime}(r_\text{inf})=\frac{\mathrm{d}}{\mathrm{d}r}\left(\frac{\mathrm{e}^{\mathrm{i}\omega r_n^*}}{r}\mathrm{e}^{v(r_n)}\right)_{r=r_\text{inf}},
    \end{split}
\end{equation}
where $r_\text{inf}$ is a large distance representing infinity (we chose $r_\text{inf}=15000M$), and
\begin{equation}
    v(r_n):=\sum_{n=1}^{n_\text{inf}}\frac{c_n}{r^n}.
\end{equation}
Here $n_\text{inf}$ stands for the cutoff imposed on the sum (we chose $n_\text{inf}=100$; in principle, the sum over $n$ should extend to infinity).

Substituting the ansatz $\phi_{\omega \ell}^{\mathrm{up}}(r_n)=(\mathrm{e}^{\mathrm{i}\omega r_n^*}/r)\mathrm{e}^{v(r_n)}$ to Eq.~\eqref{modes} leads to the condition which must be satisfied by $v(r_n)$
\begin{equation}
    r^2(r-2M)v''(r_n)+r^2(r-2M)(v'(r_n))^2+2r(M+\mathrm{i}\omega r^2)v'(r_n)-(\ell(\ell+1)r+2M)=0.
\end{equation}
Collecting the powers of $r$ and demanding that the coefficients multiplying each power vanish, we get the values of consecutive $c_n$.

%--------------------------------------------
\subsubsection{Boundary conditions for the in-modes}
%--------------------------------------------
We proceed similarly with the in-modes. We fix the boundary conditions at the horizon
\begin{equation}
    \begin{split}
        &\phi_{\omega \ell}^{\mathrm{in}}(r_\text{H})=\left.\frac{\mathrm{e}^{-\mathrm{i}\omega r_n^*}}{r}w(r_n)\right|_{r=r_\text{H}},\\
        &\phi_{\omega \ell}^{\mathrm{in}\prime}(r_\text{H})=\frac{\mathrm{d}}{\mathrm{d}r}\left(\frac{\mathrm{e}^{-\mathrm{i}\omega r_n^*}}{r}w(r_n)\right)_{r=r_\text{H}},
    \end{split}
\end{equation}
where $r_\text{H}$ represents the horizon (we chose $r_\text{H}=(20000001/10000000)M$), and
\begin{equation}
    w(r_n):=\sum_{n=0}^{n_\text{H}}b_n(r-2M)^n.
\end{equation}
Again, the sum should in principle extend to infinity, but we imposed a numerical cutoff at $n_\text{H}=200$.

Now, we substitute $\phi_{\omega \ell}^{\mathrm{up}}(r_n)=(\mathrm{e}^{-\mathrm{i}\omega r_n^*}/r)w(r_n)$ to Eq.~\eqref{modes}, and obtain the condition for the function $w(r_n)$:
\begin{equation}
    r^2(r-2M)w''(r_n)+2r(M-i\omega r^2)w'(r_n)-(\ell(\ell+1)r+2M)w(r_n)=0.
\end{equation}
Once again, we collect the powers of $r$ and solve for the coefficients $b_n$.

%--------------------------------------------
\subsubsection{Normalization}
%--------------------------------------------
Following \cite{Hodgkinson1}, we now impose normalization on the modes. It is chosen in such a way that
\begin{equation}
    \int_{-\infty}^\infty \mathrm{d}r_n^* R_{\omega\ell}R_{\omega'\ell}=2\pi\delta(\omega-\omega'),
\end{equation}
where $R_{\omega\ell}=r\Phi_{\omega\ell}$. It turns out that, to fulfill this condition, we must have
\begin{equation}
    \Phi_{\omega\ell}^{\mathrm{up}}(r_n)=B_{\omega\ell}\phi_{\omega\ell}^{\mathrm{up}}(r_n),\qquad \Phi_{\omega\ell}^{\mathrm{in}}(r_n)=B_{\omega\ell}\phi_{\omega\ell}^{\mathrm{in}}(r_n),
\end{equation}
where $B_{\omega\ell}$ is defined by
\begin{equation}
    \begin{split}
        &B_{\omega\ell}=\frac{2\mathrm{i}\omega}{W[\rho_{\omega\ell}^\mathrm{in},\rho_{\omega\ell}^\mathrm{up}]},\\
        &W[\rho_{\omega\ell}^\mathrm{in},\rho_{\omega\ell}^\mathrm{up}]=\rho_{\omega\ell}^\mathrm{in}(r_n^*)\rho_{\omega\ell}^{\mathrm{up}\prime}(r_n^*)-\rho_{\omega\ell}^\mathrm{up}(r_n^*)\rho_{\omega\ell}^{\mathrm{in}\prime}(r_n^*),\\
        &\rho_{\omega\ell}^\mathrm{up}(r_n^*)=r\phi_{\omega\ell}^{\mathrm{up}}(r_n),\qquad \rho_{\omega\ell}^\mathrm{in}(r_n^*)=r\phi_{\omega\ell}^{\mathrm{in}}(r_n).
    \end{split}
\end{equation}
The quantity $W[\rho_{\omega\ell}^\mathrm{in},\rho_{\omega\ell}^\mathrm{up}]$ (Wronskian) is constant along $r$, and can be computed at any point (we chose $r=r_H$).

\bibliography{apssamp}

\end{document}